\documentclass[aps,showpacs,preprintnumbers,amsmath,amssymb,nofootinbib]{revtex4}

\usepackage{graphicx}
\usepackage{color}

\def \be  {\begin{equation}}
\def \ee  {\end{equation}}
\def \ee  {\end{equation}}
\def \bea {\begin{eqnarray}}
\def \eea {\end{eqnarray}}

\begin{document}

\preprint{ECTP-2012-02}

\title{Effects of quantum gravity on the inflationary parameters and thermodynamics of the early universe }

\author{A.~Tawfik}
\email{a.tawfik@eng.mti.edu.eg}
\email{atawfik@cern.ch}
\affiliation{Egyptian Center for Theoretical Physics (ECTP), MTI University, Cairo, Egypt}
\affiliation{Research Center for Einstein Physics, Freie-University Berlin, Berlin, Germany}

\author{H.~Magdy}
\email{h.magdy@eng.mti.edu.eg}
\affiliation{Egyptian Center for Theoretical Physics (ECTP), MTI University, Cairo, Egypt}

\author{A. Farag Ali}
\email{ahmed.ali@fsc.bu.edu.eg}
\email{ahmed.ali@uleth.ca}
\affiliation{Physics Department, Faculty of Science, Benha University, Benha 13518, Egypt}

\date{\today}

\begin{abstract}
The effects of generalized uncertainty principle (GUP) on the  inflationary dynamics and the thermodynamics of the early universe are studied. Using the GUP approach, the tensorial and scalar density fluctuations in the inflation era are evaluated and compared with the standard case. We find a good agreement with the Wilkinson Microwave Anisotropy Probe data. Assuming that a quantum gas of scalar particles is confined within a thin layer near the apparent horizon of the Friedmann-Lemaitre-Robertson-Walker universe which satisfies the boundary condition, the number and entropy densities and the free energy arising form the quantum states are calculated using the GUP approach. A qualitative estimation for effects of the quantum gravity on all these thermodynamic quantities is introduced.

\end{abstract}

\pacs{98.80.Cq, 04.60.-m, 04.60.Bc}
\keywords{Inflation, Quantum Gravity Phenomenology}

\maketitle


\section{Introduction}

The idea that the uncertainty principle would be affected by the quantum gravity has been suggested couple decades ago \cite{maad}. Should the theories of quantum gravity, such as string theory, doubly special relativity and black hole physics be confirmed, our understanding of the basic laws and principles of physics turn to be considerably different, especially at very high energies or short distances \cite{guppapers,amati,BHGUP,garay1,hossenfelder}. Various examples can be mentioned to support this phenomena. In the context of polymer quantization, the commutation relations are given in terms of the polymer mass scale \cite{polymer}. Also, the standard  commutation relations in the quantum mechanics are conjectured to be changed or better to say generalized  at the length scales of the order of Planck's length \cite{garay1,gupps2}. Such modifications are supposed to play an essential role in the quantum gravitational corrections at very high energy \cite{qgc}. Accordingly, the standard uncertainty relation of quantum mechanics is replaced by a gravitational uncertainty relation having a minimal observable length of the order of Planck's length \cite{hossenfelder,gupp3a,gupp3b,gupp3c,gupp3d}.

The existence of a minimal length is one of the most interesting predictions of such new physics. These can be seen as the consequences of the string theory, since strings can not interact at distances smaller than their size which leads to a generalized uncertainty principle (GUP) \cite{guppapers}. Furthermore, the black hole physics suggests that the uncertainty relation should be modified near the Planck's energy scale because of measuring the photons emitted from the black hole suffers from two major errors. The first one is the error by Heisenberg classical analysis and the second one is because the black hole mass varies during the emission process and the radius of the horizon changes accordingly \cite{guppapers,BHGUP,kmm,kempf,brau,sabine}. As discussed, these newly-discovered fundamental properties of space-time would result in different  phenomenological outcomes in other physical branches \cite{bib14}. In the first part of this present work, we want to investigate the effects of GUP on the inflationary parameters in the standard inflation.

At very short distances, the holographic principle for gravity is assumed to relate the gravitational quantum theory to quantum field theory. At this short scale, the entropy of a black hole would be related to the area of the horizon \cite{entr1,entr3}. The covariant entropy bound in the Friedmann-Lemaitre-Robertson-Walker (FLRW) is found to indicate to a holographic nature in terms of temperature and entropy \cite{entr4}. The cosmological boundary can be chosen as the cosmological apparent horizon instead of the event horizon of a black hole. In light of this, we mention that the statistical (informational) entropy of a black hole can be calculated using the brick wall method \cite{entr5}. In order to avoid the divergence near the event horizon, a cutoff parameter would be utilized. Since the degrees of freedom would be dominant near horizon, the brick wall method is used to be replaced by a thin-layer model making the calculation of entropy possible \cite{entr6a,entr6b,entr6c,entr6d,entr6e,entr6f,entr6g,entr6h}. The entropy of the FLRW universe is given by time-dependent metric. The GUP approach has been used in calculating the entropy of various black holes \cite{entr7a,entr7b,entr7c,entr7d,entr7e,entr7f,entr7g,entr7h,entr7i,entr7j,entr7k,entr7l}.
The effect of GUP on the reheating phase after inflation of the universe has been studied in \cite{reheat}. The present work aims to complete this investigation by studying the effect of GUP in the inflationary era itself.
In doing this, we start from the number density arising from the quantum states in the early universe. Then, we calculate the free energy and entropy density. The idea of calculating thermodynamic quantities from quantum nature of physical systems dates back to a about one decade \cite{qentra,qentrb,qentrc,qentrd,qentre,qentrf}, where the entropy arising from mixing of the quantum states of degenerate quarks in a very simple hadronic model has been estimated and applied to different physical systems.

Some basic features of the FLRW universe are given in section \ref{sec:flrw}. The GUP Approach which will be utilized in the present work is elaborated in section \ref{sec:gup}. The whole treatment is based on the inflation era.  The consequences for the next eras of the cosmological Universe history and the recent observations on the inflation parameters are elaborated in sections \ref{eras} and \ref{inflparam}, respectively. Section \ref{sec:therm} is devoted to the second topic of the present work, the number of quantum states of the early universe.  The conclusions are listed out in section \ref{sec:conc}.


\section{The FLRW Universe}
\label{sec:flrw}

In the FLRW universe, the standard $(n+1)$-dimensional metric reads
\bea\label{eq:mtrc}
d s^2 &=& h_{ab}\, d x^a\, dx^b + r^2\, d\Omega^2_{n-1},
\eea
 where $x^a=(t,r)$ and $h_{ab}=\mathtt{diag}(-1,a^2/(1-k r^2))$. $ d\Omega^2_{n-1}$ is the line element of an $n+1$-dimensional unit sphere. $a(t)$ and $k$ are scale factor and curvature parameter, respectively. Then, the radius of the apparent horizon is given by
 \bea
 R_A &=& \left(H^2+\frac{k}{a^2}\right)^{-1/2}. \label{eq:eins0}
 \eea
It is obvious that the time evolution of the scale factor entirely depends on the background equation of state. Seeking for simplicity, we utilize \cite{zhu}
 \bea
 a(t) &=& t^{2/3 \bar{k}},
 \eea
 where $t$ is the cosmic time and $\bar{k}=1-(b c)^2/(1-c^2)$. The parameters $b$ and $c$ are free and dimensionless. Their values can be fixed by cosmological observations.
 Then, the Hubble parameter and radius of the apparent horizon read
 \bea
 H(t) &=& \frac{2}{3} \frac{1}{\bar{k}\, a^{3 \bar{k}/2}},\\
 R_A &=& \left(H \sqrt{1 + \left(\frac{3}{2}  \bar{k} \right)^{4/3  \bar{k}}\, H^{4/3  \bar{k}-2} k }\right)^{-1/2}.
 \eea
From the metric given in Eq. (\ref{eq:mtrc}) and the Einstein in non-viscous background
equations, we get
\bea
H^2 +\frac{k}{a^2} &=& \frac{8 \pi G}{3} \rho +\frac{\Lambda}{3}, \label{eq:eins1}\\
\dot H - \frac{k}{a^2} &=& -4 \pi G (\rho + p). \label{eq:eins2}
\eea
Then, the total energy $\rho$ and temperature $T$ inside the sphere of radius $R_A$ can be evaluated as follows.
 \bea
  \rho &=& \frac{\pi^{n/2}}{\Gamma\left(\frac{n}{2}\right)+1} \, \frac{n(n-1)}{16 \pi G} \, R_A^{n-1},\\
  T &=& \frac{R_A}{2 \pi} H^2 \left|1 + \frac{1}{2 H^2} \left(\dot H + \frac{k}{a^2}\right)\right|,
  \eea
where $n$ gives the dimension of the universe. From Eq. (\ref{eq:eins0}) and (\ref{eq:eins1}), it is obvious that the inverse radius of the apparent horizon is to be determined by the energy-momentum tensor i.e., matter and cosmological constant $\Lambda$. $G$ is the gravitational constant and $p$ is the pressure. Taking into consideration the viscous nature of the background geometry makes the treatment of thermodynamics of FLRW considerably complicated \cite{Tawfik:2011gh,Tawfik:2011sh,Tawfik:2011mw,Tawfik:2010pm,Tawfik:2010mb,Tawfik:2010bm,Tawfik:2010ht,Tawfik:2009nh,Tawfik:2009mk}.
For completeness, we give the cross section of particle production
  \bea
 \sigma &=& \frac{1}{M_{p}^2} \left[\frac{\rho}{M_{p}} \left(\frac{8 \Gamma\left(\frac{n}{2}\right)}{n-2}\right)\right]^{2/(n-2)},
 \eea
 where $ \Gamma$ is the gamma function and  $M_{p}$ is the Planck mass.


 \section{Tensorial and scalar density fluctuations in the inflation era}
 \label{sec:gup}

At short distances, the standard commutation relations are conjectured to be changed. In light of this, a new model of GUP was proposed \cite{advplb,Ali:2010yn,Das:2010zf}. It predicts a maximum observable momentum and a minimal measurable length. Accordingly, $[x_i, x_j]=[p_i, p_j]=0$ (via the Jacobi identity) turn to be produced.
\bea
[x_i, p_j]\hspace{-1ex} &=&\hspace{-1ex} i \hbar\hspace{-0.5ex} \left[  \delta_{ij}\hspace{-0.5ex} - \hspace{-0.5ex} \alpha \hspace{-0.5ex}  \left( p\, \delta_{ij} +
\frac{p_i p_j}{p} \right) + \alpha^2 \hspace{-0.5ex} \left( p^2\, \delta_{ij}  + 3\, p_{i}\, p_{j} \right) \hspace{-0.5ex} \right], \label{eq:alfaa}
\label{comm01}
\eea
where the parameter $\alpha = {\alpha_0}/{M_{p}c} = {\alpha_0 \ell_{p}}/{\hbar}$ and $M_{p} c^2$ stands for Planck's energy. $M_{p}$ and $\ell_{p}$ is Planck's mass and length, respectively. Apparently, Eqs. (\ref{comm01}) imply the existence of a minimum measurable length and a maximum measurable momentum
\bea
\Delta x_{min} & \approx & \alpha_0\ell_{p} \label{dxmin} , \\
\Delta p_{max} &\approx & \frac{M_{p}c}{\alpha_0} \label{dpmax},
\eea
\par\noindent
where $\Delta x \geq \Delta x_{min}$ and $\Delta p \leq \Delta p_{max}$.
Accordingly, for a particle having a distant origin and an energy scale comparable to the Planck's one, the momentum would be a subject of a  modification \cite{advplb,Ali:2010yn,Das:2010zf}
\bea
p_i &=& p_{0i} \left(1 - \alpha p_0 + 2\alpha^2 p_0^2 \right), \label{mom1} \\
 x_i &=&  x_{0i},
\eea
where $p_0^2=\sum_i p_{0i} p_{0i}$ and  $p_{0i}$ are the components of the low energy momentum. The operators $p_{0j}$ and $ x_{0i}$ satisfy the canonical commutation relation $[x_{0i}, p_{0j}] = i \hbar~\delta_{ij}$. Having the standard representation in position space, then  $p_{0i} = -i \hbar \partial/\partial{x_{0i}}$ and $x_{0i}$ would represent
the spatial coordinates operator at low energy\cite{advplb}. 

As given in \cite{Ali:2011fa} and Eq. (\ref{eq:alfaa}), the first bound for the dimensionless $\alpha_0$ is about  $\sim10^{ 17}$, which would approximately gives $\alpha\sim 10^{-2}~$GeV$^{-1}$. The other bound  of $\alpha_0$  which is $\sim10^{10}$. This lower bound means that $\alpha\sim10^{-9}~$GeV$^{-1}$. As discussed in \cite{Tawfik:2012hz}, the exact bound on $\alpha$ can be obtained by comparing with observations and experiments \cite{ref17}. It seems that the gamma rays burst would allow us to set an upper value for the GUP-charactering parameter $\alpha$.

In order to relate this with the inflation era, we define $\phi$ as the scaler field deriving the inflation in the early universe. Then, the pressure and energy density respectively read
\begin{eqnarray}
 P(\phi) &=& \frac{1}{2}\,\dot{\phi}^2 - V(\phi),    \label{6} \\
 \rho(\phi) &=& \frac{1}{2}\,\dot{\phi}^2 + V(\phi), \label{5}
\end{eqnarray}
where $V(\phi)$ is the inflation potential, which is supposed to be sufficiently flat. The main potential slow-roll parameters \cite{slowroll} are given as
\begin{eqnarray}
    \epsilon &=& \frac{{M_{p}}^2}{2}\,\left(\frac{\acute{V}(\phi)}{V(\phi)}\right)^2,\label{7}\\
    \eta &=& {M_{p}}^2\,\frac{\acute{\acute{V}}(\phi)}{V(\phi)}, \label{8}
\end{eqnarray}
where $M_{p}=m_{p}/\sqrt{8 \pi}$ is a four dimensional fundamental scale. It gives the reduced Planck's mass. The slow-roll approximations guarantee that the quantities in Eq. (\ref{7}) and (\ref{8}) are much smaller than unity.  These conditions are supposed to ensure an inflationary phase in which the expansion of the universe is accelerating.
The conformal time is given as
\begin{equation}\label{9}
    \tau = - \frac{1}{a\,H},
\end{equation}
where $a$ is the scale factor and $H=\dot{a}/a$ is the Hubble parameter.

In order to distinguish from the curvature parameter $k$, which is widely used in literature, let us denote the wave number by $j$. Here, $j$ is assumed to give the {\it comoving} momentum. It seems to be $\tau$-dependent and can be expressed is terms of the {\it physical} momentum $P$
\begin{equation}\label{10}
    j = a P = - \frac{P}{\tau\,H}.
\end{equation}
In the GUP approach, the momentum is subject of modification, $j \longrightarrow j(1-\alpha\,j)$. Accordingly, the modification in the comic scale $a$ reads
\begin{equation}\label{11}
    a = \frac{j(1-\alpha\,j)}{P}.
\end{equation}
Then, in the presence of minimal length cutoff, the scalar spectral index is given by
\begin{eqnarray}
    n_s  &=& \frac{d\,\ln p_s}{d\,\ln j(1-\alpha\,j)} +1
    \simeq (1-\alpha\,j)\,\frac{d\,\ln p_s}{d\,\ln j}+1. \label{12}
\end{eqnarray}
 where $p_s$ is the amplitude of the scalar density perturbation i.e., the scalar density fluctuations. Recent observations on the inflation parameters are elaborated to section \ref{inflparam}. Due to the modified commutators, a change in $H$ is likely expected. This can be realized using slow-roll parameters. In the standard case, the spectral index can be expressed in these quantities \cite{tensorRa1},
 \bea \label{eq:nss}
 n_s &=& 1+2\, \eta - 6\, \epsilon,
 \eea
 where $\eta$ and $\epsilon$ are given in Eqs (\ref{7}) and (\ref{8}).
 Finally the ''running'' of the spectral index is given by
 \bea \label{nr1}
 n_r &=& \frac{d\, n_s}{d\, \ln j} = 16\, \epsilon\, \eta - 24\, \epsilon^2 - 2\, \zeta,
 \eea
 where
 \bea
 \zeta &=& M_{p} \frac{\acute{V}(\phi)\, \acute{\acute{V}}(\phi)}{V^2(\phi)},
 \eea
 is another slow-roll parameter.
 At the horizon crossing epoch, the derivative of Hubble parameter $H$ with respect to $j$ leads to \cite{slowroll,dHdK}
\begin{equation}\label{15}
    \frac{dH}{dj} = - \frac{\epsilon\,H}{j}.
\end{equation}
Therefore, when changing $j$ into $j(1-\alpha j)$, we get an approximative expression for $H$ as a function of  the modified momentum
\begin{equation}\label{17}
    H \simeq j^{-\epsilon} e^{\epsilon\,\alpha\,j}.
\end{equation}
It is obvious that  GUP seems to enhance the Hubble parameter so that $H(\alpha=0)/H(\alpha\neq 0)<1$.

One of the main consequences of inflation is the generation of primordial cosmological perturbations \cite{tensorialscalarD} and the production of long wavelength gravitational waves (tensor perturbations). Therefore, the tensorial density perturbations (gravitational waves) produced during the inflation era seem to serve as an important tool helping in discriminating among different types of inflationary models \cite{tensorialscalarB}. These perturbations typically give a much smaller contribution to the cosmic microwave background (CMB) radiation anisotropy than the inflationary adiabatic scalar perturbations \cite{tensorialscalarC}.

The tensorial and scalar density fluctuations are given as
\begin{eqnarray}
  p_t &=& \left(\frac{H}{2\pi}\right)^{2} \left[1-\frac{H}{\Lambda}\,\sin\left(\frac{2\Lambda}{H}\right)\right]
   = \left(\frac{k^{-\epsilon} e^{\epsilon\,\alpha\,k}}{2\pi}\right)^{2} \left[1-\frac {k^{\epsilon-1} e^{-\epsilon\,\alpha\,k}}{a}\,\sin\left(\frac{2}{a k^{1-\epsilon} e^{\epsilon\,\alpha\,k}}\right)\right], \\
  p_s &=& \left(\frac{H}{\dot{\phi}}\right)^2\left(\frac{H}{2\pi}\right)^{2} \left[1-\frac{H}{\Lambda}\,\sin\left(\frac{2\Lambda}{H}\right)\right]
   = \left(\frac{H}{\dot{\phi}}\right)^2\left(\frac{k^{-\epsilon} e^{\epsilon\,\alpha\,k}}{2\pi}\right)^{2} \left[1-\frac {k^{\epsilon-1} e^{-\epsilon\,\alpha\,k}}{a}\,\sin\left(\frac{2}{a k^{1-\epsilon} e^{\epsilon\,\alpha\,k}}\right)\right],
\end{eqnarray}
respectively. Then, the ratio of tensor-to-scalar fluctuations, $p_t/p_s$,  \cite{tensorRa1,tensorialscalar,tensorialscalarB} reads
\begin{equation}\label{18}
    \frac{p_t}{p_s} = \left(\frac{\dot{\phi}}{H}\right)^{2}. 
    \end{equation}
In the standard case, this ratio is assumed to linearly depend on the inflation slow-roll parameters \cite{tensorRa1}
\bea \label{eq:cal1}
 \frac{p_t}{p_s} = {\cal O}(\epsilon).
 \eea

It is apparent that Eq. (\ref{17}) gives an estimation for $H$ in terms of the wave number $j$. To estimate $\dot\phi$, we start with the equation of motion for the scalar field, the Klein-Gordon equation,
\bea \label{eq:ddphi1}
\ddot\phi + 3 H \dot\phi + \partial_{\phi} V(\phi) &=& 0.
\eea
The $\dot\phi$-term has the same role as that of the friction term in classical mechanics. In order to get inflation from a scalar field, we assume that Eq. (\ref{eq:ddphi1}) is valid for a very flat potential leading to neglecting its acceleration i.e., neglecting the first term. Some inflationary models introduce the slow-roll parameter $\eta_H=-\ddot\phi/H\dot\phi=-\ddot H/2 H \dot H$. Therefore, the requirement to neglect $\ddot\phi$ is equivalent to guarantee that $\eta_H<<1$.
\bea
\dot\phi &=& - \frac{1}{3 H}\, \partial_{\phi} V(\phi),
\eea
where the potential itself is model-dependent, for example, $V(\phi) =
M_{p} \exp[- \sqrt{2/H_0 p}\, \phi]$ \cite{effectss}. In our model, 
\bea
\dot{\phi} = \left(\frac{\sqrt{2\epsilon}\,V}{M_{p}\,H}\right)^2.
\eea
Then, the tensor-to-scalar fluctuations ratio reads
\bea
\frac{p_t}{p_s} &=&  \left(\frac{\sqrt{2}\,V}{M_{p}}\, \frac{\sqrt{\epsilon}}{j^{-2\epsilon} e^{2\epsilon\,\alpha\,j  }}\right)^2.
\eea

\begin{figure}[htb]
\includegraphics[width=5.5cm,angle=-90]{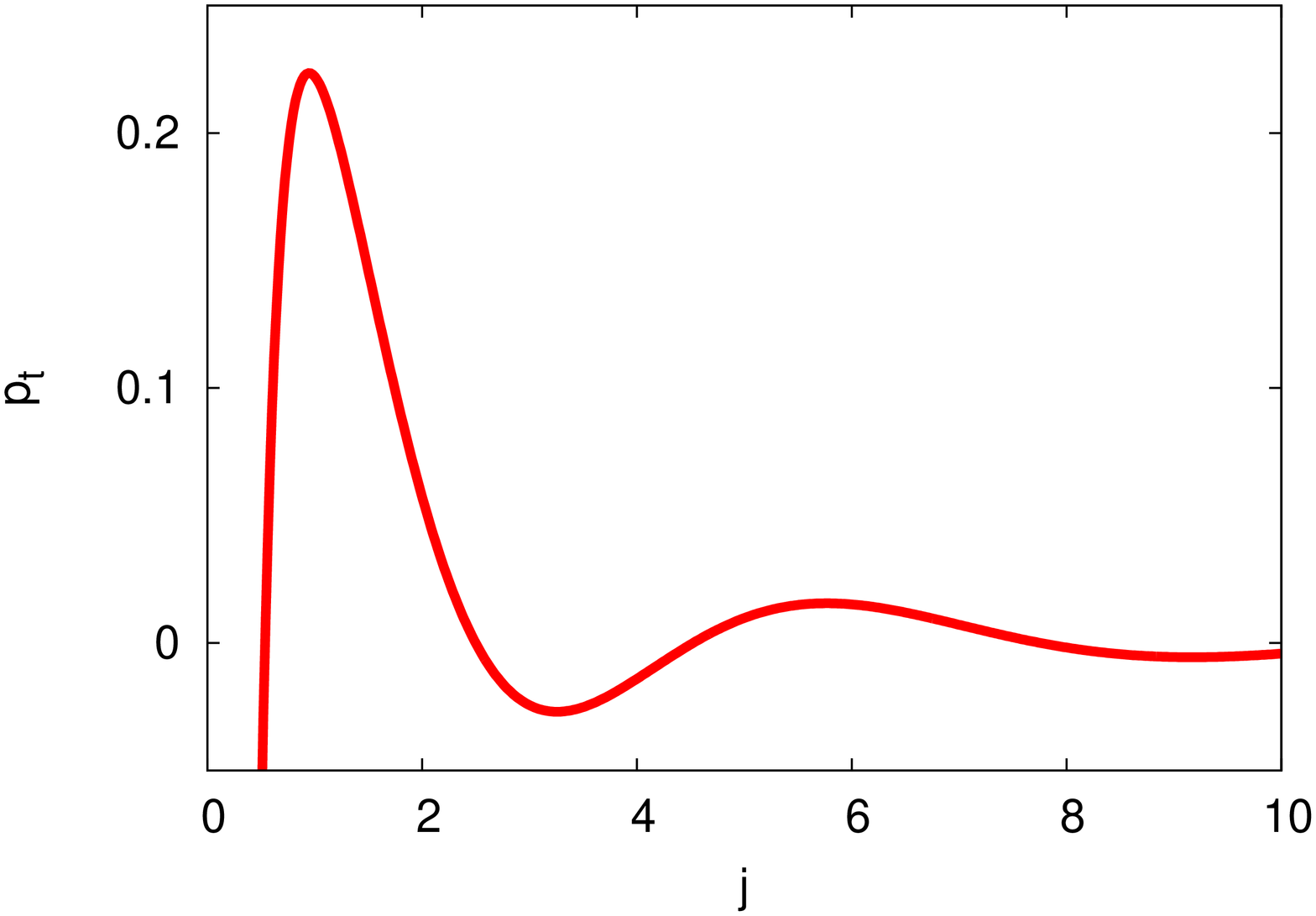}
\includegraphics[width=5.5cm,angle=-90]{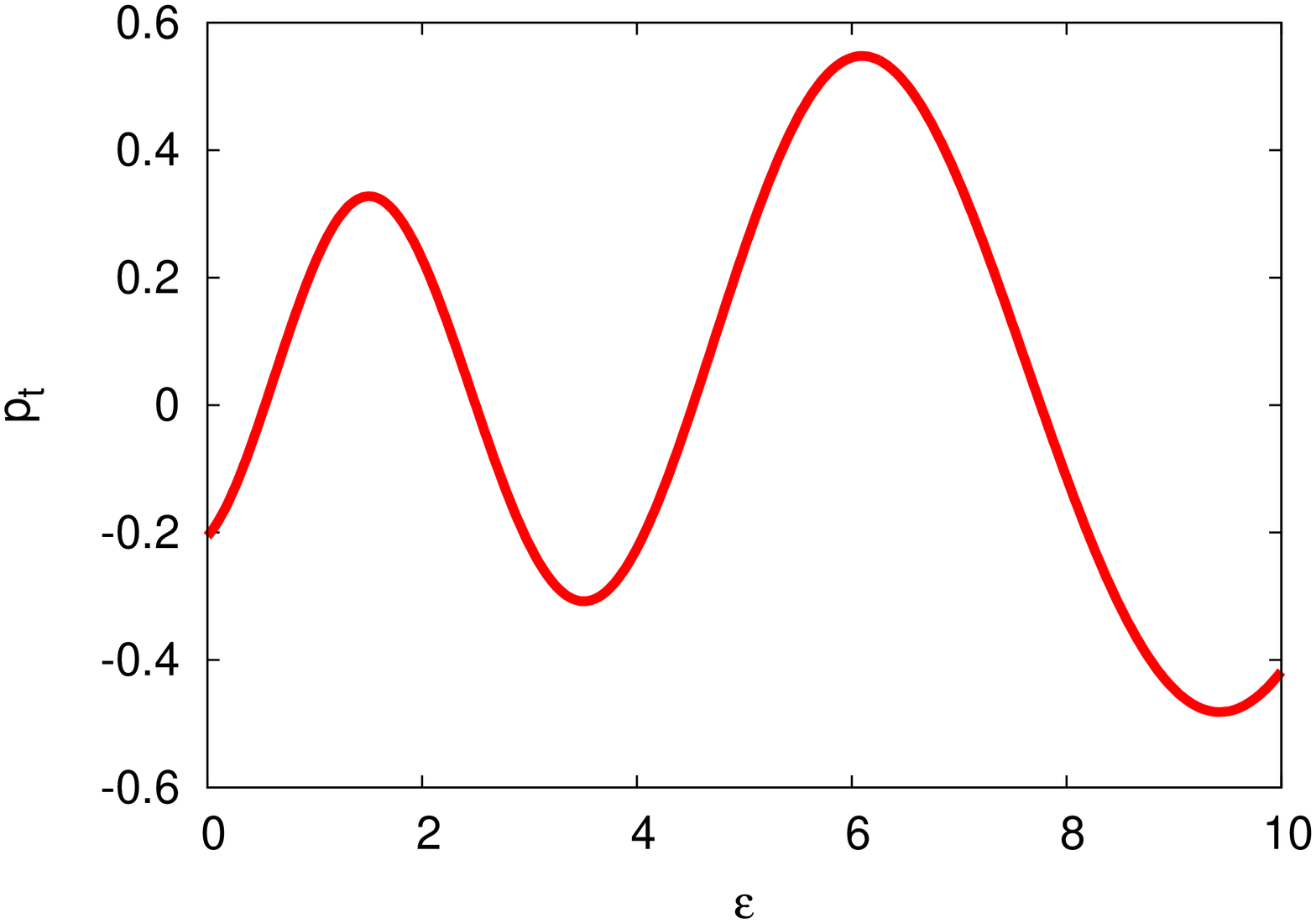}
\caption{The tensorial density fluctuations $p_t$ is given in dependence on the wave number $j$ (left panel) and on the slow-roll parameter $\epsilon$ (right panel). The parameter $\alpha$ is kept constant, $\alpha=10^{-2}~$GeV$^{-1}$ (lower bound). It is assumed the $\sqrt{2}\,  V/M_{p}$ remains constant, (nearly unity). These two assumptions set the physical scale.}
\label{fig:Pt1}
\end{figure}

Fig. \ref{fig:Pt1} gives the tensorial density fluctuations $p_t$ in dependence on the wave number $j$ (left panel) and on the slow-roll parameter $\epsilon$ (right panel). In both graphs, $\alpha$ is kept constant, $\alpha=10^{-2}~$GeV$^{-1}$ i.e., the upper bound is utilized. Also, it is assumed the potential is nearly of the order of the reduced mass $M_{p}$ i.e., $\sqrt{2} V/M_{p}\sim 1$. It is obvious that $p_t$ diverges to negative value at low $j$. Increasing $j$ brings $p_t$ to positive values. After reaching a maximum value, it  decreases almost exponentially and simultaneously oscillates around the abscissa. The amplitude of oscillation drastically decreases with increasing $j$. The right panel shows that $p_t(\epsilon)$ oscillates around the abscissa. Here, the amplitude of the oscillation raises with increasing $\epsilon$. The oscillation can be detected essentially in the CMB spectrum quantizing the primordial residuals of the quantum gravity effects.

\begin{figure}[htb]
\includegraphics[width=5.5cm,angle=-90]{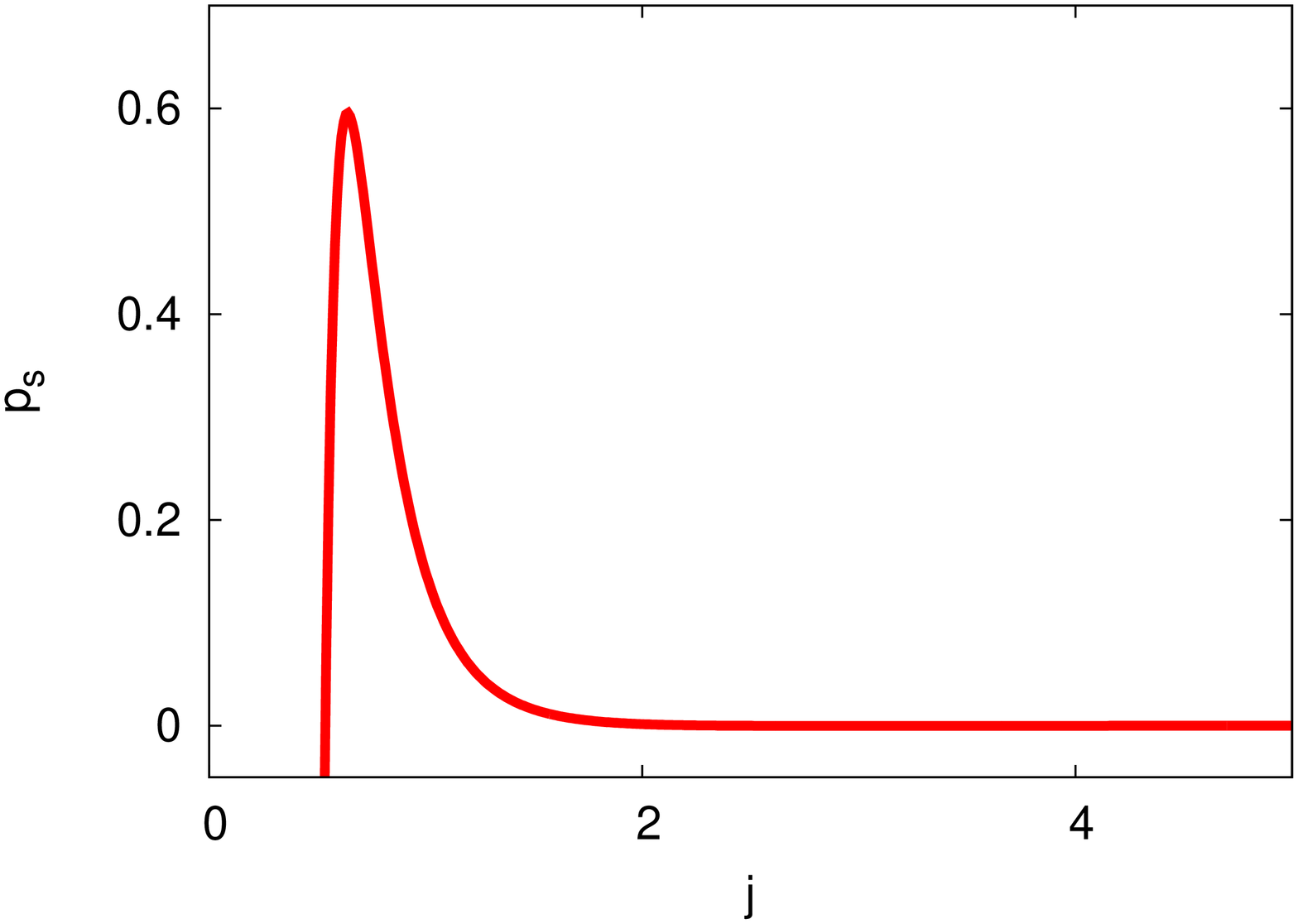}
\includegraphics[width=5.5cm,angle=-90]{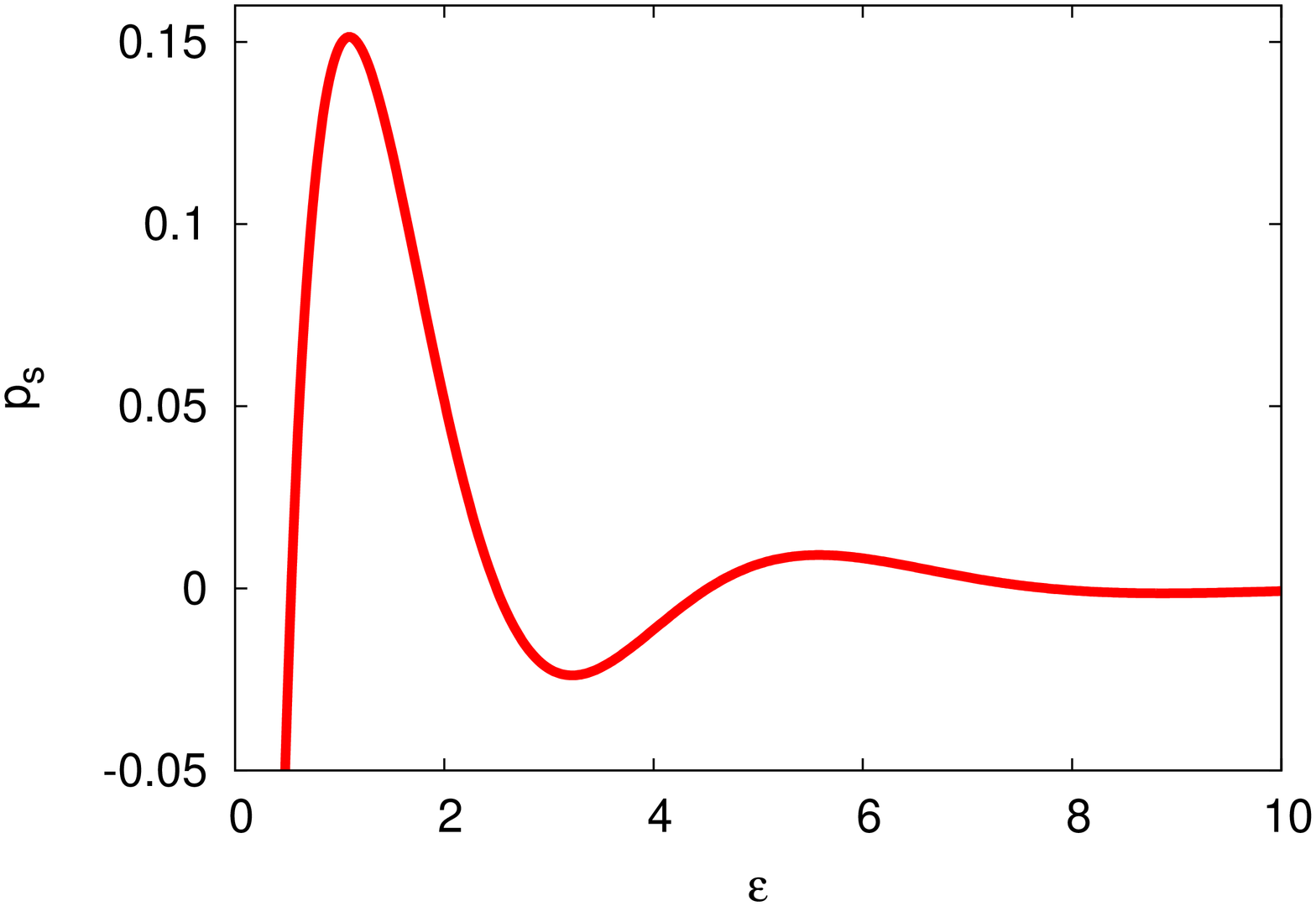}
\caption{The scalar density fluctuations $p_s$  is given in dependence on $j$ (left panel) and on slow-roll parameter $\epsilon$ (right panel). $\alpha$ and $\sqrt{2} V/M_{p}$ have the same values as in Fig. \ref{fig:Pt1}. They set the physical scale.}
\label{fig:Ps1}
\end{figure}

Fig. \ref{fig:Ps1} refers to nearly the same behavior as that of the dependence of scalar density fluctuations $p_s$  on the wave number $j$ and $\epsilon$. It is apparent that $p_s$ diverges to negative value at low $j$. Increasing $j$ brings $p_s$ to positive values. But after reaching a maximum value, it decreases almost exponentially. Nevertheless its values keep their positive sign. The oscillation of $p_s(\epsilon)$ is also observed. Here, $p_s(\epsilon)$ behaves almost similar to $p_t(k)$. After reaching a maximum value, it almost exponentially decreases and simultaneously oscillates around the abscissa. The amplitude of oscillation drastically decreases with increasing $\epsilon$.

\begin{figure}[htb]
\includegraphics[width=7.cm,angle=-90]{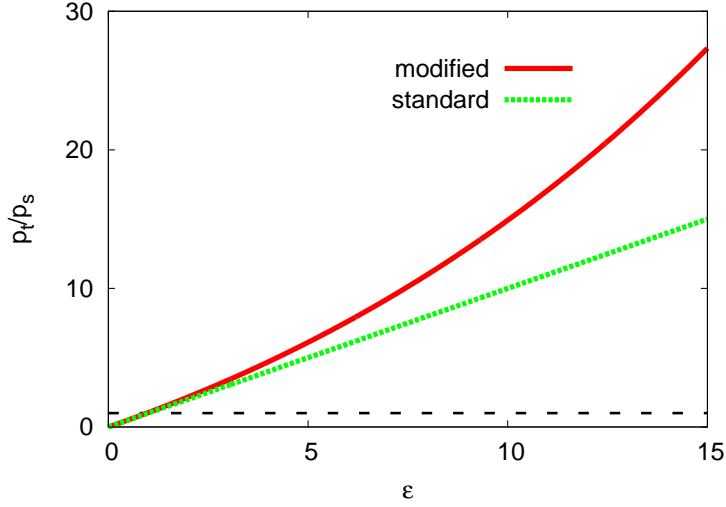}
\caption{The dependence of the ratio $p_t/p_s$ on the slow-roll parameter $\epsilon$ is given in ''standard'' and ''modified'' cases. The GUP parameter $\alpha$ (in ''modified'' case) and $\sqrt{2} V/M_{p}$ have the same values as in Fig. \ref{fig:Pt1} and therefore the physical scale is defined. The horizontal dashed line represents constant ratio $p_t/p_s$.}
\label{fig:PtPs1}
\end{figure}

Fig. \ref{fig:PtPs1} gives the ratio $p_t/p_s$ in dependence on $\epsilon$ in two cases. The first case, the ''standard'' one, is given by solid curve. The second case, the ''modified'' case, is given by dashed curve. The latter is characterized by finite $\alpha$, while in the earlier case, $\alpha$ vanishes. Compared to the ''standard'' case, there is a considerable increase in the values of  $p_t/p_s$ with raising $\epsilon$.
For the ''modifiied'' case i.e., upper bound of $\alpha=10^{-2}$ GeV$^{-1}$, the best fit results in an exponential function
\bea \label{eq:cal2}
\frac{p_t}{p_s} &=& \mu \, \epsilon^{\nu},
\eea
where $\mu=0.875\pm0.023$ and $\nu=1.217\pm0.014$. All these quantities are given in natural units. For the ''standard'' case, the results can be fitted by
\bea \label{eq:cal3}
\frac{p_t}{p_s} &=&  \epsilon,
\eea
which agrees very well with Eq. (\ref{eq:cal1}). The difference between Eqs. (\ref{eq:cal2}) and (\ref{eq:cal3}) is coming from the factor in the denominator reflecting the correction due to the GUP approach.

As discussed above, the CMB results and other astrophysical observations strongly make constrains on the standard cosmological parameters such as the Hubble parameter $H$,
baryon density $n_b$ and even age of the universe \cite{refff0,refff1}. It turns to be necessary to have constrains on the power spectrum of the primordial fluctuations \cite{refff2}. This is achievable through the spectral index. From Eq. (\ref{12}), the scalar spectral index at $\sqrt{2} V/M=1$ reads
{\small
\bea
n_s &=& 1+ \left\{4 e^{-6 j \alpha  \epsilon } j^{6 \epsilon } \pi ^2 (1-j \alpha ) \right. \nonumber \\
&& \epsilon  \left[-\frac{3}{2 \pi ^2} e^{6 j \alpha  \epsilon } j^{-6 \epsilon }\, \left(1-\frac{e^{-j \alpha  \epsilon } j^{-1+\epsilon
}}{a} \sin \left(\frac{2 e^{-j \alpha  \epsilon } j^{-1+\epsilon }}{a}\right)\right)+\right. \nonumber \\
&& \frac{3}{2 \pi ^2} e^{6 j \alpha  \epsilon } j^{1-6 \epsilon }\, \alpha  \left(1-\frac{e^{-j \alpha  \epsilon } j^{-1+\epsilon }}{a} \sin
\left(\frac{2 e^{-j \alpha  \epsilon } j^{-1+\epsilon }}{a}\right)\right)+  \nonumber \\
&& \frac{1}{4 \pi ^2\, \epsilon }e^{6 j \alpha  \epsilon } j^{-6 \epsilon }\left(-\frac{1}{a}e^{-j \alpha  \epsilon } j^{-1+\epsilon } \left(\frac{2
e^{-j \alpha  \epsilon } j^{-1+\epsilon } (-1+\epsilon )}{a}-\frac{2 e^{-j \alpha  \epsilon } j^{\epsilon } \alpha  \epsilon }{a}\right) \cos \left(\frac{2
e^{-j \alpha  \epsilon } j^{-1+\epsilon }}{a}\right)-\right. \nonumber \\
&& \left.\left.\left.\frac{e^{-j \alpha  \epsilon } j^{-1+\epsilon }}{a} (-1+\epsilon ) \sin \left(\frac{2 e^{-j \alpha  \epsilon } j^{-1+\epsilon
}}{a}\right)+\frac{e^{-j \alpha  \epsilon } j^{\epsilon } \alpha  \epsilon }{a} \sin \left(\frac{2 e^{-j \alpha  \epsilon } j^{-1+\epsilon }}{a}\right)\right)\right]\right\}/ \nonumber \\
&& \left[1-\frac{e^{-j \alpha  \epsilon } j^{-1+\epsilon }}{a} \sin \left(\frac{2 e^{-j \alpha  \epsilon } j^{-1+\epsilon }}{a}\right)\right].
\eea
}

\begin{figure}[htb]
\includegraphics[width=8.cm]{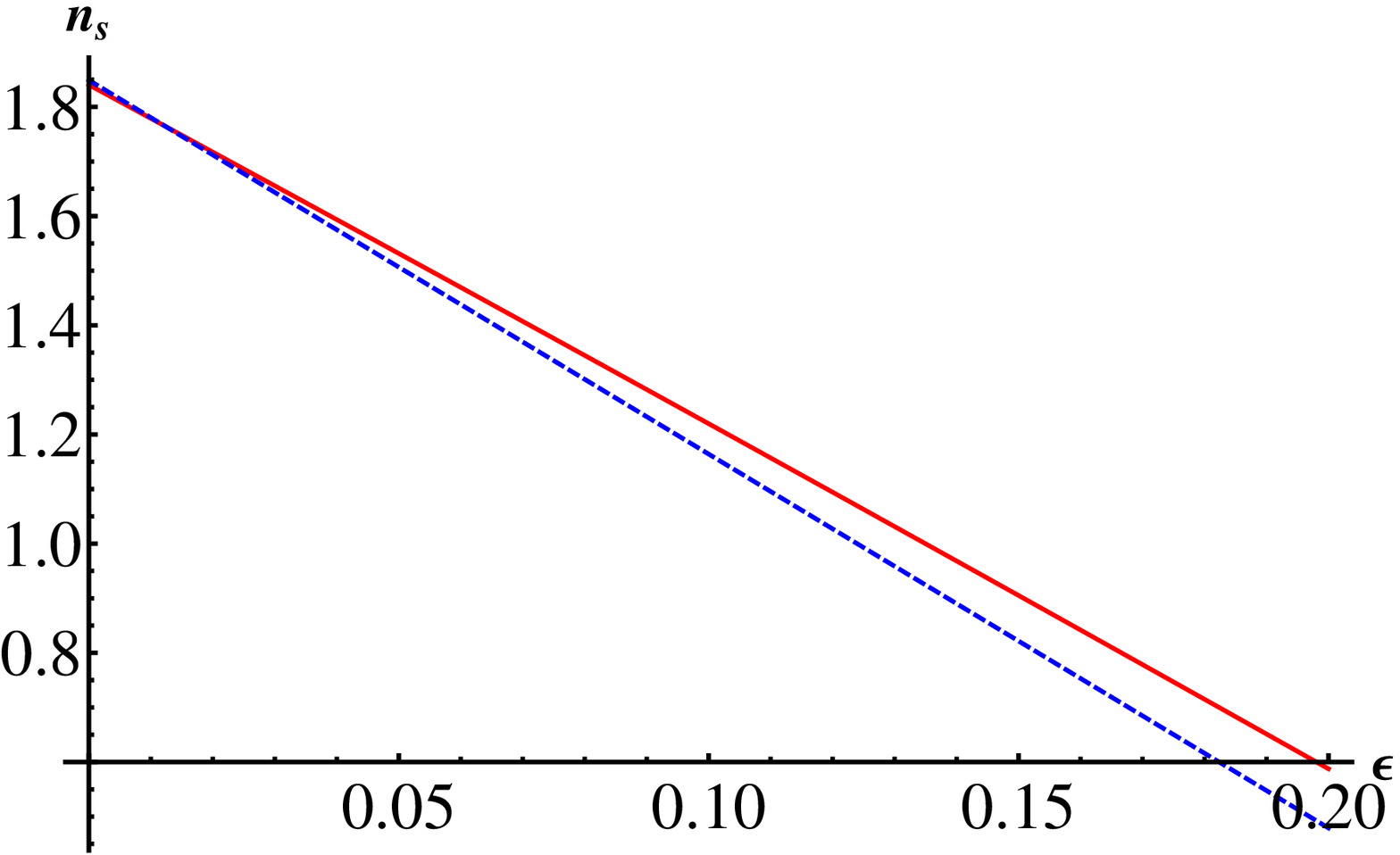}
\includegraphics[width=8.cm]{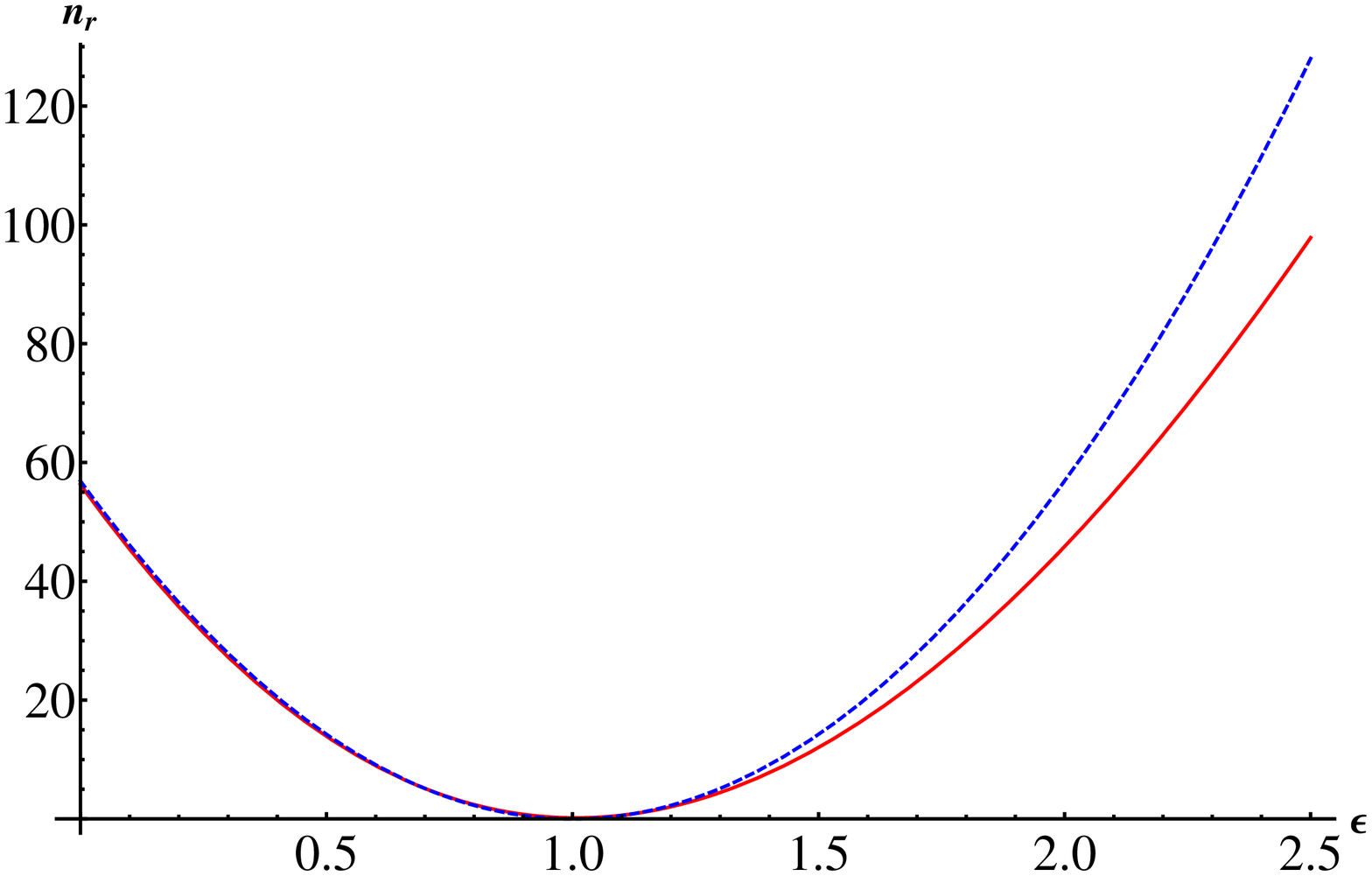}
\caption{Left panel: the spectral index $n_s$ is given in dependence on $\epsilon$, where $j$ and $a$ are kept constant (equal $1$). The ''running'' of $n_s$ is shown in the right panel. The solid curves represent the results from the modified momentum $j\rightarrow j(1-\alpha j)$ i.e., applying the GUP approach. The dashed curves represent the standard case (unchanged momentum) i.e., $\alpha=0$.  All these quantities are given in natural units.}
\label{fig:NsNr}
\end{figure}

The ''running'' of the spectral index $n_s$ is defined by Eq. (\ref{nr1}). The results of $n_r=d\, n_s/d\, \ln\, j$ are depicted in the right panel of Fig. \ref{fig:NsNr}. Early analysis of the Wilkinson Microwave Anisotropy Probe (WMAP) data \cite{wmap1,wmap2} indicates that $n_r = -0.03\pm 0.018$. As noticed in \cite{wmap2}, such analysis may require modification, as their statistical significance seems to be questionable. On the other hand, it is indicated that the spectral index quantity $n_s-1$ seems to run from positive values on long length scales to negative values on short length scales. This is also noticed in left panel of Fig. \ref{fig:NsNr}, where we draw $n_s$ vs. $\omega$. Such a coincident observation can be seen as an obvious evidence that our model agrees well with the WMAP-data. Recent WMAP analysis shows that $n_s=0.97\pm0.017$ \cite{wmap3}. The importance of such agreement would be the firm prediction of inflationary cosmology through the consistency relation between scalar and tensor spectra. The physics at the Planck's scale is conjectured to modify the consistency relation considerably. It also leads to the running of the spectral index. For modes that are larger than the current horizon, the tensor spectral index is positive \cite{runnn}.

\section{Consequences for the next eras of the cosmological Universe history}
\label{eras}

In describing the primordial power spectrum, almost all inflation models utilize three independent parameters. The first one is the amplitude of the scalar fluctuations. The second one is the tensor-to-scalar ratio $n_r$. The third one is the scalar spectral index $n_s$. All of these parameters are observationally measurable. They allow the connection between the high-energy physics and the observational cosmology, in particular CMB.

The dependence of tensor-to-scalar ratio $p_t/p_s$ on $\epsilon$ is drawn in Fig. \ref{fig:PtPs1}.  The ''modified'' momentum characterized by finite $\alpha$ and reflecting the quantum gravity effects, shows a considerable increase with raising $\epsilon$. Accordingly, the best fit results in
\bea \label{eq:cal2b}
\left.\frac{p_t}{p_s}\right|_{qc} &=& \mu \, \epsilon^{\nu},
\eea
where the subscript stands for quantum gravity. The ''standard'' case can be fitted  by
\bea \label{eq:cal3b}
\left.\frac{p_t}{p_s}\right|_{s} &=&  \epsilon.
\eea
The relation between Eqs. (\ref{eq:cal2b}) and (\ref{eq:cal3b}) can be given as
\bea
\left.\frac{p_t}{p_s}\right|_{qc} &=& \left(\frac{\mu}{\left.\frac{p_t}{p_s}\right|_{s}}\right)^{\nu},
\eea
where the values of the fitting parameters $\mu$ and $\nu$ are given in Eq. (\ref{eq:cal2}).

The dependence of $n_s$ on $\epsilon$ is presented in the left panel of Fig. \ref{fig:NsNr}, while the dependence of its ''running'',  Eq. (\ref{nr1}), is illustrated in the right panel. Including quantum gravity effects keeps the linear dependence of  $n_s(\epsilon)$ unchanged, but makes it slower than in the standard case in which the momentum remains unchanged. Increasing $\epsilon$ leads to an increase in the difference between modified and unmodified momentum. The running $n_s$ is not affected by quantum gravity at $\epsilon<1$. At higher $\epsilon$ values, $n_r$ in modified momentum gets slower than the one in standard case.

The spectral index $n_s$ describes the initial density ripples in the Universe. If $n_s$ is small, the ripples with longer wavelengths are strong, and vice versa. This has the effect of raising the CMB power spectrum on one side and lowering it on the other. $n_s$  is like a fingerprint of the very beginning of the universe in that first trillionth of a second after the Big Bang called Inflation. The way of distributing matter during the initial expansion reflects the nature of the energy field controlling the inflation. The current observations on $n_s$ are in agreement with inflation's prediction of a nearly scale-invariant power spectrum, corresponding to a slowly rolling inflation field and a slowly varying Hubble parameter during inflation. Based on Eq. (\ref{17}),  GUP seems to enhance the Hubble parameter so that $H(\alpha=0)<H(\alpha\neq 0)$.

\section{Recent Observations on the Inflation Parameters}
\label{inflparam}

As introduced in Ref. \cite{obsr1}, the observational inflation seems to predict a stochastic background of gravitational waves over a broad range of frequencies. They are connected with the cosmic microwave background (CMB) measurements, which in turn are accessible directly with gravitational-wave detectors, like NASA’s Big-Bang Observer (BBO) \cite{obsr2}. The observations of BBO are connected to CMB constraints to the amplitude and tensor spectral index of the inflationary gravitational-wave background (IGWB) for different inflationary models. Furthermore, the results obtained in the WMAP third-year data release are connected with the analysis introduced in Ref. \cite{obsr2}.

It has been noted that when $n_s\neq 1$ the amplitude of the IGWB is significant. This would be apparent from the dependence of spectral index $n_s$ on the slow-roll parameters $\epsilon$ and $\eta$ given by Eq. (\ref{eq:nss})  and from the tensor-scalar ratio $r$. In order to infer a value for $r$ given the indication that $1-n_s\approx 0.05$, one has to suppose some natural relationship between $\epsilon$ and $\eta$.  Recent measurements indicate that $n_s<1$ indicating  a significant amplitude for the gravitational-wave background produced by inflation. An extension through the inclusion of  IGWB accessible to direct observation in additional to the inclusion of  the amplitude of  IGWB accessible to the observations of the polarization of CMB has been reported \cite{obsr2}. An upper limit to $r$ and a precise measurement of $n_s$ characterize the curvature of the inflation potential,
\bea \label{eq:reta}
r &=& \frac{8}{3} \left(1-n_s+2\, \eta\right).
\eea
At constant $n_s$, Eq. (\ref{eq:reta}) describes a $r$-$\eta$ plane, from which an upper limit to $r$ can be deduced. For $n_s$ ranging from $0.94$ to $0.96$ an upper limit of $r < 0.1$ implies that the potential would have a negative curvature which would have important implications for inflationary model building. The Planck satellite is expected to attain $0.5\%$ in a determination of $n_s$ at a fiducial value $n_s=0.957$ \cite{obsr3}. This would then translate into a lower bound for Coleman-Weinberg inflation, for instance, $r>0.0046$.


\section{Number of Quantum States  in the inflation era}
\label{sec:therm}

Based on the general uncertainty principle with the minimal length, the statistical (informational) entropy of the FLRW universe described by time-dependent metric is calculated in this section. In section \ref{sec:flrw}, the FLRW cosmology is briefly reviewed. The minimal length related to the Plank scale can be related to the area of the cosmological apparent horizon. The latter would be in turn be related to $\alpha$. Interestingly, such a relation is conjectured to characterize the black holes, where its entropy is proportional to the area of its horizon \cite{entr1,entr3}. Instead of the event horizon of a black hole, the cosmological boundaries can be chosen to be identical with the apparent  cosmological horizon. It is assumed that the universe will be in locally thermodynamic equilibrium state. Similar to the black hole, the degrees of freedom of a field can be dominant near horizon.

\begin{figure}[htb]
\includegraphics[width=5.5cm]{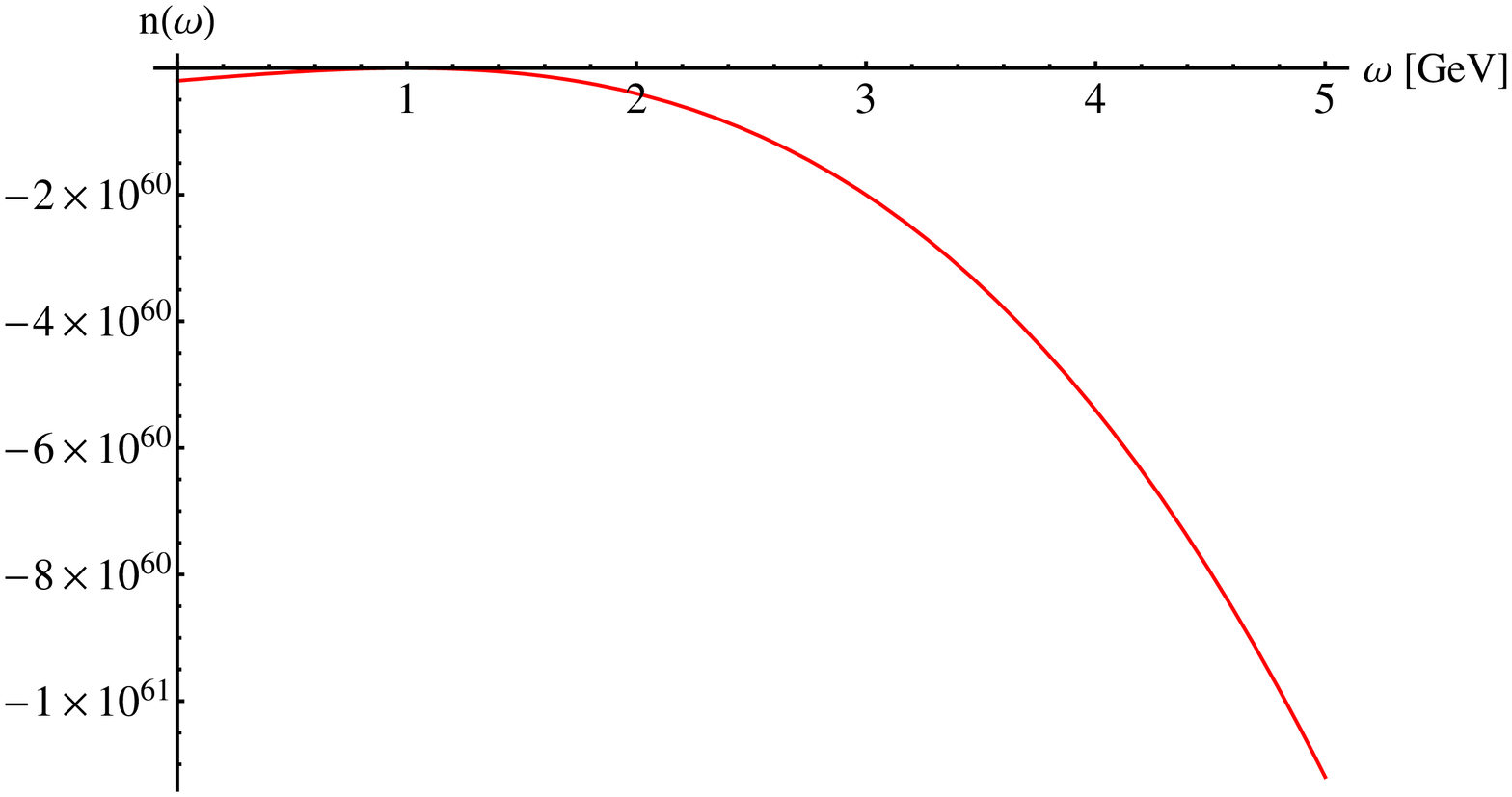}
\includegraphics[width=5.5cm]{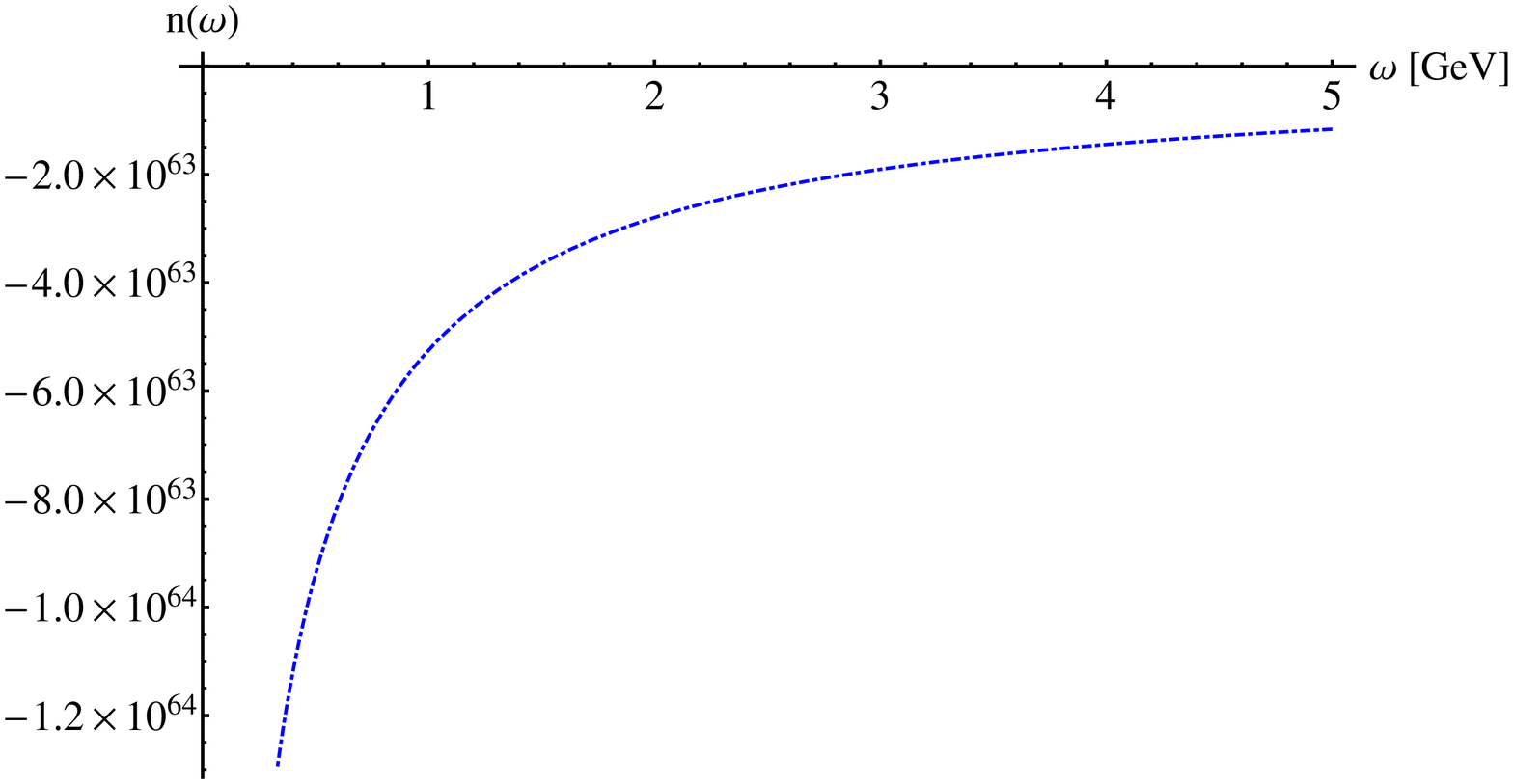}
\includegraphics[width=5.5cm]{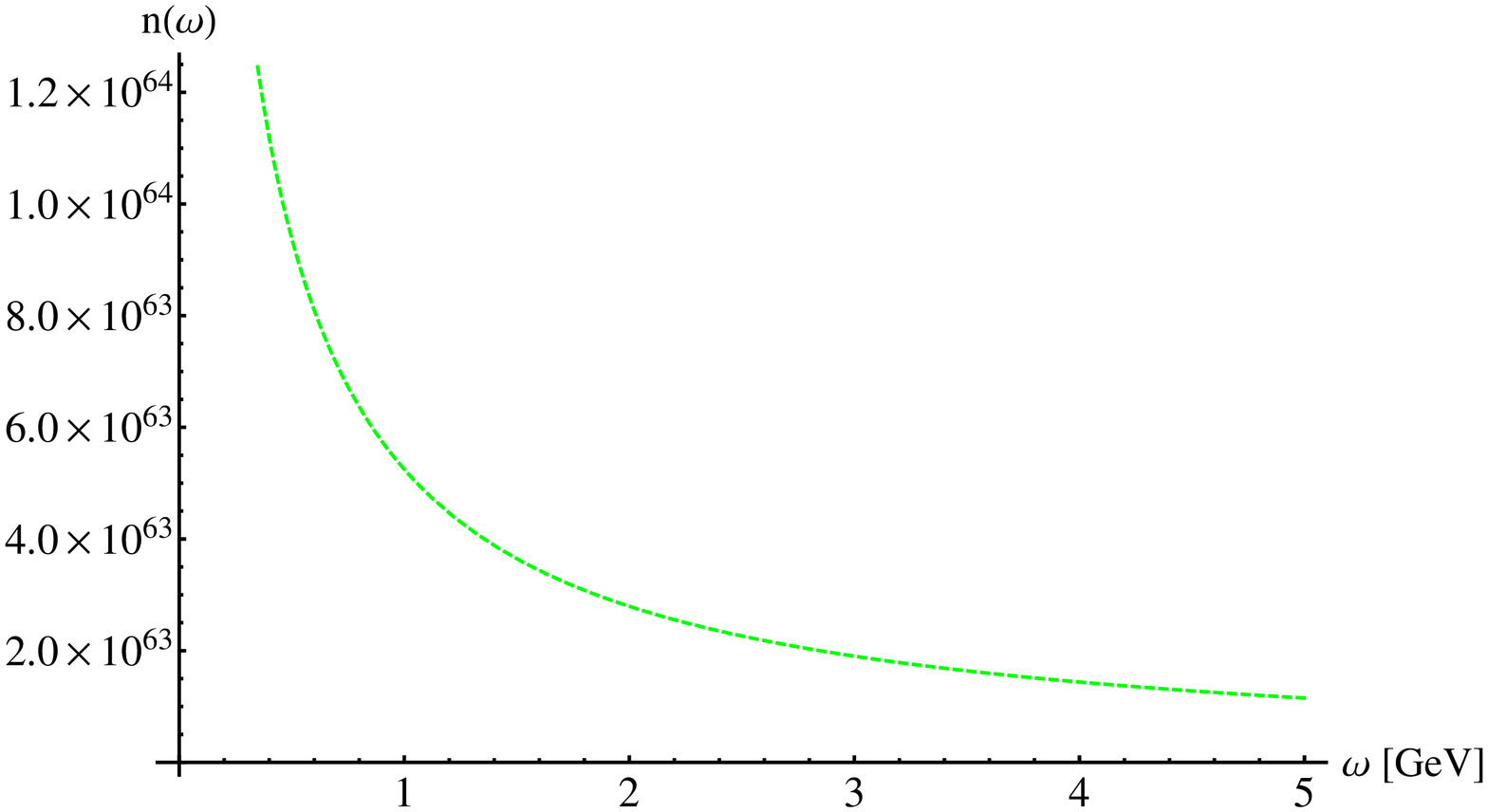}
\caption{In  natural units, the number density of the quantum states in the inflation era is given in dependence on the energy of the scalar field $\omega$. Left panel shows the results at the upper bound of $\alpha$. The results at the lower bound of $\alpha$ are given in the middle panel. The difference between the two bounds is given in the right panel.}
\label{fig:Nn1}
\end{figure}

When assuming that a quantum gas of scalar particles is confined within a thin layer near
the apparent horizon of the FLRW universe satisfying the boundary condition, the number of quantum states can be calculated using  the GUP approach. In calculating this, we take into consideration a potential change in the phase space \cite{phases}.
\begin{eqnarray} \label{eq:nomega1}
n(\omega) &=& \frac{1}{(2 \pi)^3} \int dR\, d\theta\, d\phi\; \frac{dp_{R}\, dp_{\theta}\, dp_{\phi}}{(1-\alpha P)^4} = \frac{1}{2 \pi^2} \int \, dR \int \frac{P^2-P_R^2}{(1-\alpha P)^4} dP_R,
\end{eqnarray}
where $\omega$ and $P_R$ are the energy and momentum of the scalar field, respectively. $R$ represents the spacial dimension of the layer of interest, where the number of quantum state is to be estimated. As given above, a locally equilibrium system is assumed in which the temperature of thermal radiation is slowly varying near the horizon, so that the temperature is approximately proportional to the apparent horizon, $T\propto 1/R$.
Using natural units $c=\hbar=k_B=1$, then Eq. (\ref{eq:nomega1}) leads to
{\footnotesize
\begin{eqnarray}
n(\omega) &=&  \frac{1}{2 \pi} \int  \left\{ -\frac{1}{B}\left(3 f C \alpha ^2 \omega \right) \right.  \nonumber \\
& & \hspace*{0cm} \left. + \frac{1}{A B} \left[3 \left(f \left(25+18 \alpha ^2 F +\alpha ^4 F\right) + H^2 R^2 \alpha ^2 \omega ^2 \left(-18+\alpha ^2 \left(2 \omega ^2+\mu ^2 \left(-2+\omega ^2\right)\right)\right)\right) \right] \right. \nonumber \\
& & \hspace*{0cm} \left. -\frac{1}{A^2 B} \left[12 \left(f \left(3+4 \alpha ^2 F +\alpha ^4 F^2\right)+H^2 R^2 \alpha ^2 \omega ^2 \left(-4+\alpha ^2 \left(2 \omega ^2+\mu ^2 \left(-2+\omega ^2\right)\right)\right)\right)\right] \right. \nonumber \\
& & \hspace*{0cm} \left. +\frac{1}{A^3 B} \left[8 \left(f \left(1+\alpha ^2 F\right)^2+H^2 R^2 \alpha ^2 \omega^2 \left(-2+\alpha ^2 \left(2 \omega ^2+\mu ^2 \left(-2+\omega ^2\right)\right)\right)\right) \right]  \right. \nonumber \\
& & \hspace*{0cm} \left. + \frac{4 \alpha  f D}{A^3 B}   \left[15+\alpha ^2 \left(43 \mu ^2+(40 C-43 \omega ) \omega \right) \right. \right.  \\
& & \hspace*{1.6cm} \left.\left. +  \alpha ^4 \left(41 \mu ^4+2 \mu ^2 (37 C-41 \omega ) \omega +\omega^2 \left(33 C^2-74 C \omega +41 \omega ^2\right)\right) \right. \right.  \nonumber\\
& & \hspace*{1.6cm} \left.\left. +  \alpha ^6 \left(13 \mu ^6+\mu ^4 (34 C-39 \omega ) \omega +
\mu ^2 \omega ^2 \left(27 C^2-68 C \omega +39 \omega ^2\right)+\omega^3 \left(6 H^3 P^3 R^3-27 C^2 \omega +34 C \omega ^2-13 \omega ^3\right)\right)\right]\right.  \nonumber \\
& & \hspace*{0cm} \left.  - \frac{4 \alpha  D}{3 H R \alpha ^4 \omega A^3} \left[3+8 \alpha ^2 \left(\mu ^2+(C-\omega ) \omega \right) \right. \right. \nonumber \\
& & \hspace*{2.2cm} \left.\left. +\alpha ^4 \left(\mu ^4 \left(5+2 \omega ^2\right)+\omega ^2 \left(9 C^2-14 C \omega +5 \omega ^2\right)+2 \mu ^2 \omega  \left(-\omega  \left(5+\omega ^2\right)+C \left(7+\omega ^2\right)\right)\right)\right] \right. \nonumber \\
& & \hspace*{0cm} \left.  -\frac{1}{B}\left[12 \left(-H^2 R^2 \alpha ^2 \omega ^2+f \left(5+\alpha ^2 F\right)\right)\mathtt{atanh}(\alpha D)\right] \right. \nonumber \\
& & \hspace*{0cm} \left.  -\frac{1}{B}\left[6 \left(H^2 R^2 \alpha ^2 \omega ^2 - f \left(5+\alpha ^2 F\right)\right)\ln(A)\right] \right\} \, dR \nonumber
\end{eqnarray}
}
where $\mu$ is the mass of the field of interest,
\hbox{$A = 1 + \alpha^2 \left[\mu^2 + (C - \omega) \omega\right]$},
\hbox{$B = 3 H^3 R^3 \alpha^6 \omega^3$},
\hbox{$C = H P R$},
\hbox{$D = \sqrt{-\mu ^2+\omega  (-C+\omega )}$},
\hbox{$F =  \mu ^2-\omega^2$} and $f=1-R^2/R_A^2$ is a function of the comoving time $t$ and dimension of the layer $R$ with $R_A=1/(H^2+k/a^2)^{1/2}$.
For $D=1$ and because of the small value of $\alpha$, all high orders of $\alpha$ are disregarded. Based on these assumptions the number density of the quantum states is given as
{\small
\begin{eqnarray}
n(\omega) &=& \frac{1}{3 f^2 H^3 \alpha ^6 \omega ^3} \left\{ \frac{1}{2 R^2}\left[ -6 H^5 R^5 \alpha ^7 \omega ^3 \left(H R \left(2+3 \omega  (1+\omega )^2\right)-2 K (3+\omega  (3+\omega ))\right)-\right.\right. \nonumber \\
&& 2 f^2 H R \alpha ^2 \omega K \left(-3+8 \alpha  \left(1+\alpha ^2 F \right) \left(20+17 \alpha ^2 F\right)\right)-  \nonumber \\
&& 2 f H^3 R^3 \alpha ^5 \omega  \left(K \left(4 \omega  (33+4 \omega )+\alpha ^2 \left(-\omega
^2 (9+\omega  (117+43 \omega ))+\mu ^2 \left(9+\omega  \left(117+43 \omega +4 \omega ^3\right)\right)\right)\right)-\right. \nonumber \\
&&
\left.2 H R \omega  \left(33+\omega  (8+33 \omega )+\alpha ^2 \left(-\omega ^2 (27+4 \omega  (8+9 \omega ))+\mu ^2 (27+2 \omega  (16+\omega
 (18+\omega )))\right)\right)\right)+ \nonumber \\
&&
\left. f^3 \left(-47+\alpha  \left(-60-\alpha F \left(22+\alpha  \left(172+\alpha  F
\left(-1+164 \alpha +52 \alpha ^3 F \right)\right)\right)\right)\right)\right]+ \label{eq:noemaga1} \\
&& \frac{f^2}{R^2} \left[ 6 \mathtt{atanh}[\alpha ] \left(f \left(5+\alpha ^2 F \right)+2 H^2 R^2 \alpha ^2 \omega ^2
\mathtt{ln}(R)\right)+H^2 R^2 \alpha ^2 \left\{\right. \right. \nonumber \\
&&\left.  \hspace*{4mm}  \omega  \left(3-22 \omega +\alpha  \left(-4 (40+3 \omega )+\alpha  \left(2 \mu ^2 \omega -\left(2+\mu ^2\right) \omega
^3 - 4 \alpha F (74+41 \omega) -8 \alpha ^3 F \left(F (17+16 \omega ) + \mu ^2 \omega ^3 \right)\right)\right)\right) \ln(R)+\right. \nonumber \\
&&  \hspace*{4mm} \left(-3 \omega ^2+160 \alpha  \omega ^2+12 \alpha ^3  \omega  F  (11+26 \omega )+\alpha ^5 F \left(\omega ^2 (9-\omega  (99+173 \omega ))+\mu ^2 \left(-9+\omega  \left(99+173 \omega +4 \omega ^3\right)\right)\right)\right) \nonumber \\
&& \hspace*{3mm} \left.\left.\left.\mathtt{ln}\left(2 \left(H R \omega + K \right)\right)\right\}\right]\right\}, \nonumber
\end{eqnarray} }
where $K=\sqrt{-f \mu ^2+\left(f+H^2 R^2\right)\omega ^2}$. The results of $n(\omega)$ are shown in Fig. \ref{fig:Nn1}. Left panel shows the results at the upper bound of $\alpha$. The lower bound is given in the middle panel, which would be nearly identical with the standard case, where $\alpha=0$. Subtracting the results from the lower bound from the results from the upper bound seems to result in the affect of the quantum gravity. These results are illustrated in the left panel.

In Boltzmann limit, the entropy can be directly derived from the free energy is given by $-\int  n(\omega)/(\exp{\beta \omega}-1) d \omega$. The entropy reads $\beta^2 \int \omega n(\omega)/(4 \sinh^2(\beta \omega/2)) d \omega$. The resulting expression is given in Appendix \ref{app:b}. The results are shown in Fig. \ref{fig:Ss1}.  In performing these calculations, we set  $H=10^{-25}$ GeV, $R=10^{34}$ GeV$^{-1}$, $f=1$, $\mu=1$ GeV and $\beta=2.4 \times 10^{-4}$ GeV. It has been shown that the thermodynamic first law is fulfilled. When the cosmological constant $\Lambda$ becomes dominant compared to other forms of matter, the entropy is found to be satisfied $S=A/4G$, where $A$ is the area of the apparent horizon. Again, we distinguish between upper and lower bound of $\alpha$. The lower bound is assumed to be very identical with the standard case.

\section{Conclusions}
\label{sec:conc}
We have studied the effects of GUP on the inflationary dynamics and thermodynamics of the early universe. In the first part of this work, we introduce an evaluation for the tensorial and scalar density fluctuations in the inflation era. Furthermore, comparing with standard case, the case in which the effects of quantum gravity are excluded, we give an estimation for the GUP on all these parameters. The tensorial  $p_t$ and scalar density fluctuations $p_s$ are given in dependence on the wave number $j$ and on the slow-roll parameter $\epsilon$. For a systematic comparison, the parameter $\alpha$ is is kept constant, $\alpha=10^{-2}~$GeV$^{-1}$ i.e., the upper bound is utilized. Also, it is assumed the $\sqrt{2} V/M_{p}\sim 1$. We noticed that $p_t$ diverges to negative value at low $j$. Increasing $j$ brings $p_t$ to positive values. After reaching a maximum value, it almost exponentially decreases and simultaneously oscillates around the abscissa. The amplitude of oscillation drastically decreases with increasing $j$. Also, $p_t(\epsilon)$ is founf to oscillate around the abscissa. Here, the amplitude of the oscillation raises with increasing $\epsilon$. The oscillation can be detected essentially in the CMB spectrum quantizing the primordial residuals of the quantum gravity effects.

The spectral scalar index $n_s$ is defined by scalar index. The running of this essential parameter in conjectures to shed light on its scaling.  The WMAP data indicates that the spectral index quantity $n_s-1$ seems to run from positive values on long length scales to negative values on short length scales \cite{wmapns1}. This behavior is confirmed in our calculations. The importance of such agreement would be the firm prediction of inflationary cosmology through the consistency relation between scalar and tensor spectra. The Planck scale physics is conjectured to modify the consistency relation considerably. It also leads to the running of the spectral index, as seen in our calculations. For modes that are larger than the current horizon, the tensor spectral index is positive. The limitation to the apparent cosmological horizon has been discussed in the present work

Assuming that a quantum gas of scalar particles is confined within a thin layer near the apparent horizon of the FLRW universe which satisfies the boundary condition, the number and entropy densities and the free energy arising form the quantum states are calculated using the GUP approach. When taking into consideration the quantum gravity i.e., applying the GUP approach, a qualitative estimation for the effects of the quantum gravity on all these thermodynamic quantities is introduced.

\section*{Acknowledgments}
The research of AT has been partly supported by the German--Egyptian Scientific Projects (GESP ID: 1378). AT likes to thank Prof. Antonino Zichichi for his kind invitation to attend the twenty-ninth World Laboratory Meeting at the ''Ettore Majorana Foundation and Centre for Scientific Culture'' in Erice-Italy, where the present work is completed. The research of AFA is supported by Benha University. The authors gratefully thank the anonymous referee for useful comments and suggestions which helped to improve the paper.


\appendix

\section{Entropy and free energy}
\label{app:b}
At temperature $T=1/\beta$, the entropy can be deduced from  the number of quantum states, Eq. (\ref{eq:noemaga1}), \cite{imprt}

\begin{figure}[htb]
\includegraphics[width=5.5cm]{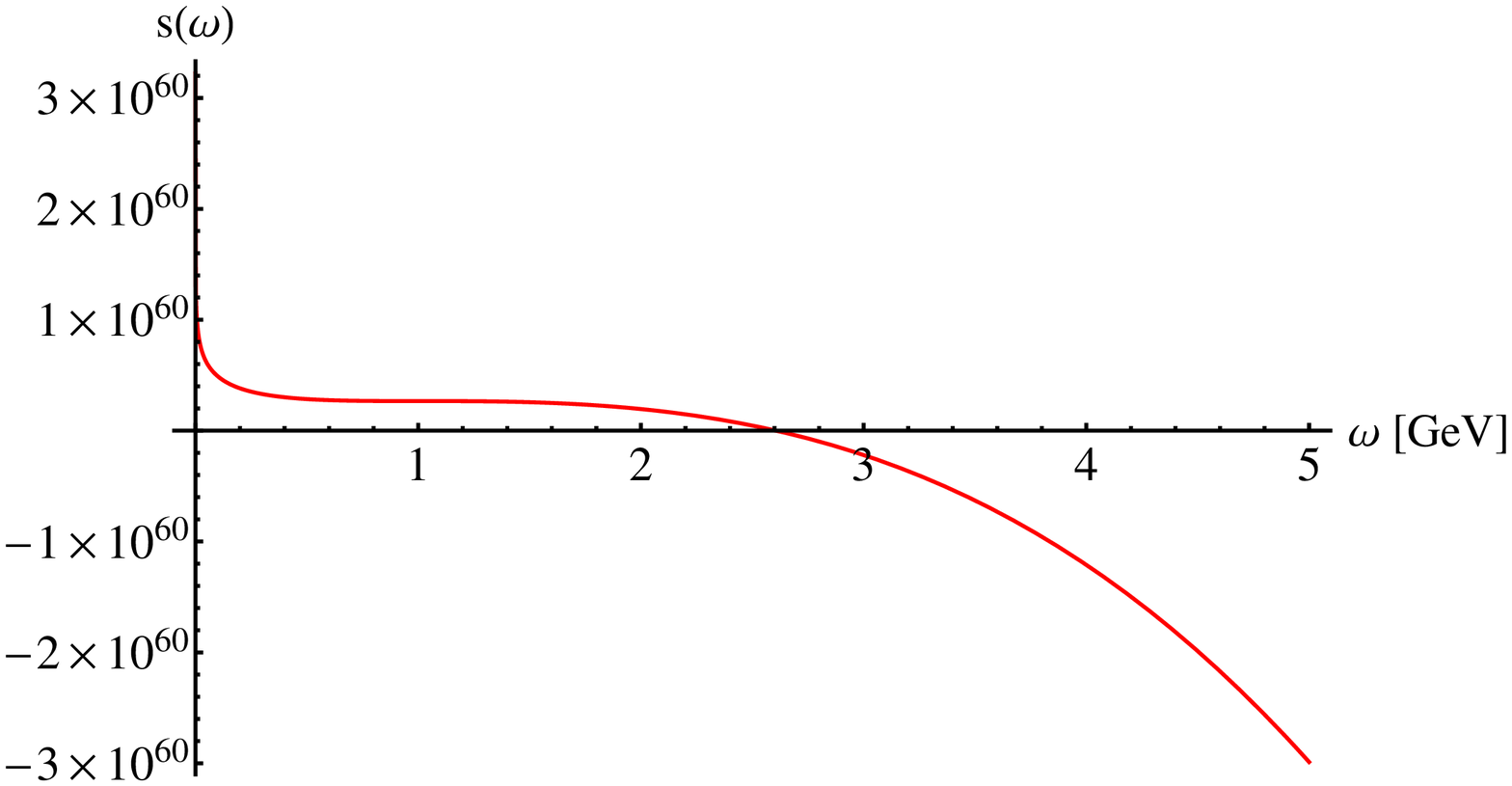}
\includegraphics[width=5.5cm]{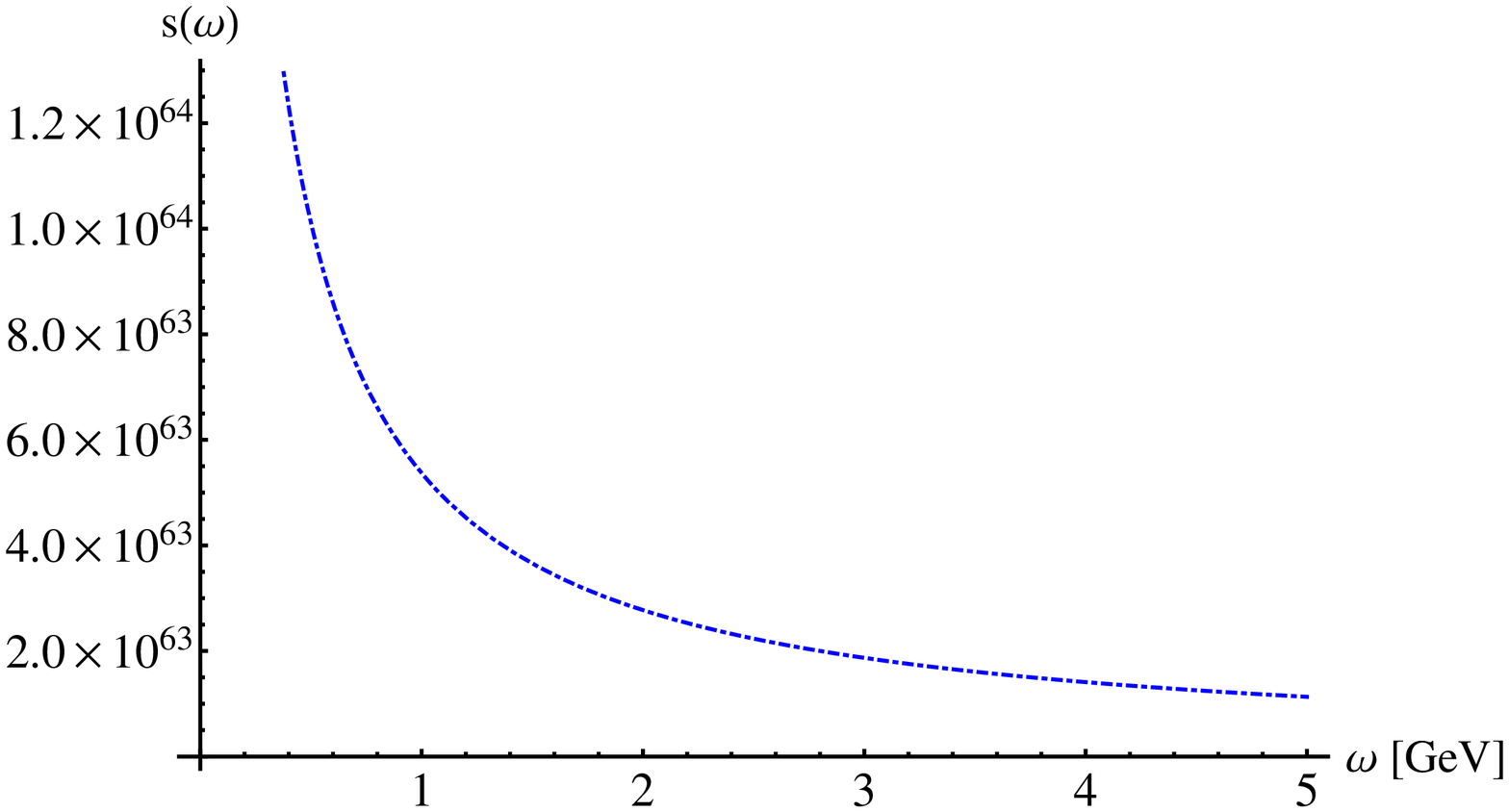}
\includegraphics[width=5.5cm]{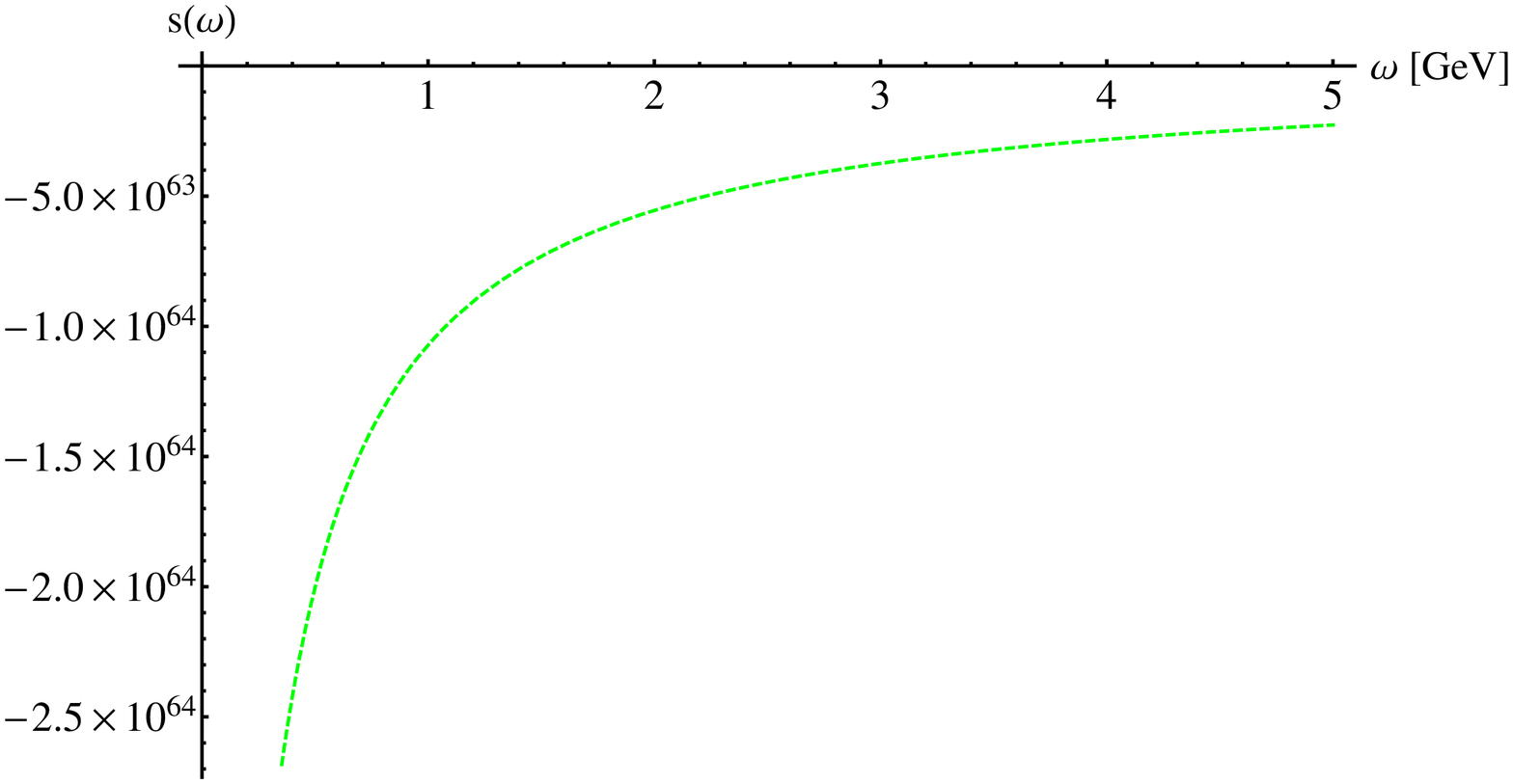}
\caption{In  natural units, the entropy density of the quantum states in the inflation era is given in dependence on the energy of the scalar field $\omega$. As in Fig. \ref{fig:Nn1}, the left and middle panels show the results at the upper and lower bounds of $\alpha$, respectively. The difference between upper- and lower-bound-results is shown in the right panel.}
\label{fig:Ss1}
\end{figure}

{\small
\begin{eqnarray}
s(\omega) &=& \beta^2 \int  \frac{\omega n(\omega)}{4 \sinh^2(\beta \omega/2)}d \omega \nonumber \\
&=&\frac{1}{18 H^3 R^2 \alpha ^6} \left\{ \frac{1}{\omega ^3}\left(-12 \mathtt{atanh}(\alpha) \left(f \left(5+\alpha ^2 \left(\mu ^2-3 \omega ^2\right)\right)+6 H^2 R^2 \alpha^2 \omega ^2 \ln(R)\right)+\right.\right. \nonumber \\
&&  H^2 R^2 \alpha ^2 \omega  \ln(R) [ -9+132 \omega + 2 \alpha
\left(240+36 \omega +\alpha  \left(-6 \mu ^2 \omega -3 \left(2+\mu ^2\right) \omega ^3+12 \alpha  \left(41 \omega ^3+\mu ^2 (37+41 \omega )\right) - \right. \right. \nonumber \\
&&\hspace*{5mm} \left.  \left. 4 \alpha ^3 \left(\omega ^4 (51+32 \omega) - 2 \mu ^2 \omega ^3 \left(96+\omega ^2\right) + \mu ^4 \left(-51-96 \omega +6 \omega ^3\right) \right)\right)\right)+ \nonumber \\
& & \hspace*{5mm} \left.\left. 48 \alpha ^3 \left(37+34 \alpha ^2 \mu ^2\right) \omega ^2 \ln(\omega)\right]\right)+ \nonumber \\
& & \frac{1}{f^2}\left(\frac{f^3}{\omega^3} \left(47+\alpha  \left(60+2 \alpha  (11+86 \alpha) \mu^2+\alpha ^3 (-1+164 \alpha ) \mu^4+52 \alpha^6 \mu^6\right)\right)-\right. \nonumber \\
& &  \frac{1}{\omega }6 f \alpha ^2 \left(6 H^4 R^4 \alpha ^3 \left(11+9 \alpha ^2 \mu ^2\right)+f^2 \left(11+\alpha  \left(86+\alpha  (-1+164 \alpha
) \mu ^2+78 \alpha ^4 \mu ^4\right)\right)\right)- \nonumber \\
&& 3 \alpha ^4 \left(18 H^6 R^6 \alpha ^3+f^3 \left(-1+164 \alpha +156 \alpha ^3 \mu ^2\right)-12 f H^4 R^4 \alpha  \left(11+3 \alpha ^2 \left(-3+4
\mu ^2\right)\right)\right) \omega - \nonumber \\
&&  6 H^4 R^4 \alpha ^7 \left(9 H^2 R^2-2 f \left(-16+\mu ^2\right)\right) \omega ^2 + 2 \left(26 f^3-72 f H^4 R^4-9 H^6 R^6\right) \alpha ^7 \omega ^3+\frac{H R \alpha ^2 K}{\left(f+H^2 R^2\right) \omega ^2} \nonumber \\
&& \left(6 H^6 R^6 \alpha ^5 \omega ^2 (18+\omega  (9+2 \omega ))+f^3 \left(-9+480 \alpha +888 \alpha ^3 \left(\mu ^2+2 \omega ^2\right)+136 \alpha
^5 \left(3 \mu ^4+14 \mu ^2 \omega ^2-2 \omega ^4\right)\right)+\right. \nonumber \\
&& f H^4 R^4 \alpha ^3 \left(24 (33-4 \omega ) \omega +\alpha ^2 \left(\mu ^2 \left(27+702 \omega -270 \omega ^2-8 \omega ^4\right)+\omega ^2 (162+\omega
 (405+98 \omega ))\right)\right)+ \nonumber \\
 && \left.f^2 H^2 R^2 \left(-9+480 \alpha +24 \alpha ^3 \left(37 \mu ^2+\omega  (33+70 \omega )\right)+\alpha ^5 \left(3 \omega ^2 (18+(117-62 \omega
) \omega )+8 \mu ^4 \left(51+\omega ^2\right) + \right.\right.\right. \nonumber \\
&& \left.\left.\left. \mu ^2 \left(27+2 \omega  \left(351+644 \omega -4 \omega ^3\right)\right)\right)\right)\right)+12 H^4 R^4 \alpha ^5 \left(-3 H^2 R^2 \alpha ^2+8 f \left(1+4 \alpha ^2 \mu ^2\right)\right) \ln(\omega)+ \nonumber \\
&& \frac{1}{\mu }3 i \sqrt{f} H R \alpha ^2 \left(f \left(f+H^2 R^2\right) (-3+160 \alpha )+3 \alpha ^3 \left(296 f^2+15 H^4 R^4 \alpha ^2+f H^2
R^2 \left(88+9 \alpha ^2\right)\right) \mu ^2+\right. \nonumber \\
&& \left. 10 f \left(68 f+5 H^2 R^2\right) \alpha ^5 \mu ^4\right)\ln\left[\left.\left(2 i \sqrt{f} \mu -2 K \right)\right/\right. \nonumber \\
&& \left.\left(f H R \alpha ^2 \left(f \left(f+H^2 R^2\right) (-3+160 \alpha )+3 \alpha ^3 \left(296 f^2+15 H^4 R^4 \alpha ^2+f H^2 R^2 \left(88+9\alpha ^2\right)\right) \mu ^2+\right.\right.\right. \nonumber \\
& & \left.\left.\left.\left.\left. 10 f \left(68 f+5 H^2 R^2\right) \alpha^5 \mu^4\right) \omega \right)\right] - \right.\right. \nonumber \\
&& \left.\left. \frac{9 f H^3 R^3 \alpha ^5}{\sqrt{f+H^2 R^2}} \left(88 \left(f+H^2 R^2\right)+3 \left(39 f+28 H^2 R^2\right) \alpha^2 \mu^2\right)
\ln\left[2 \left(f \omega +H^2 R^2 \omega +\sqrt{f+H^2 R^2} K \right)\right]\right)\right\}.
\end{eqnarray}
}
We notice that the expression contains complex terms. The results at upper and lower bounds of $\alpha$ are given in Fig. \ref{fig:Ss1} (shown in the left and middle panels, respectively). The difference between $s(\omega)$ at upper and lower bound of $\alpha$ is given in the right panel. Only real values are drawn.

\begin{figure}[htb]
\includegraphics[width=5.5cm]{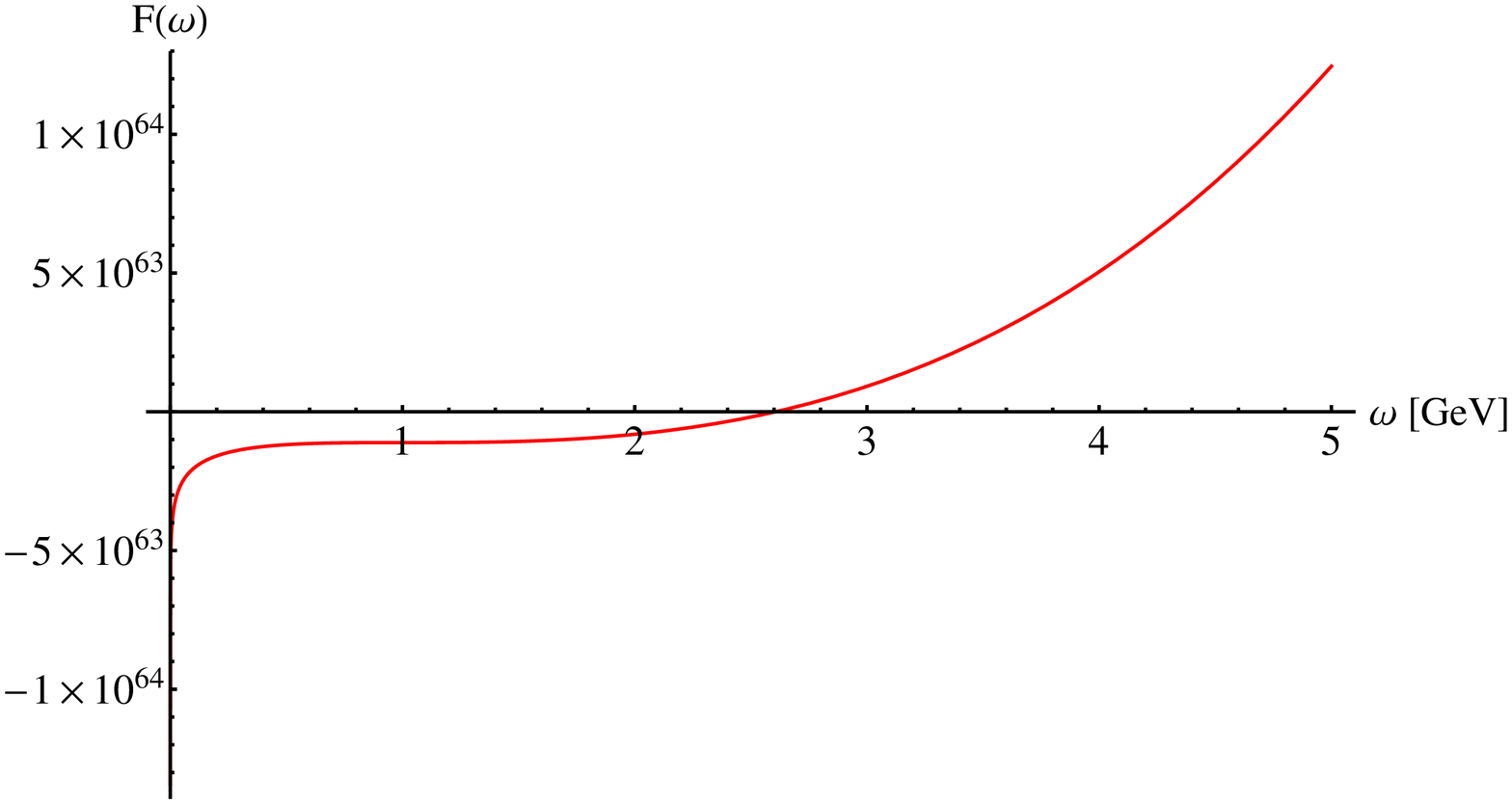}
\includegraphics[width=5.5cm]{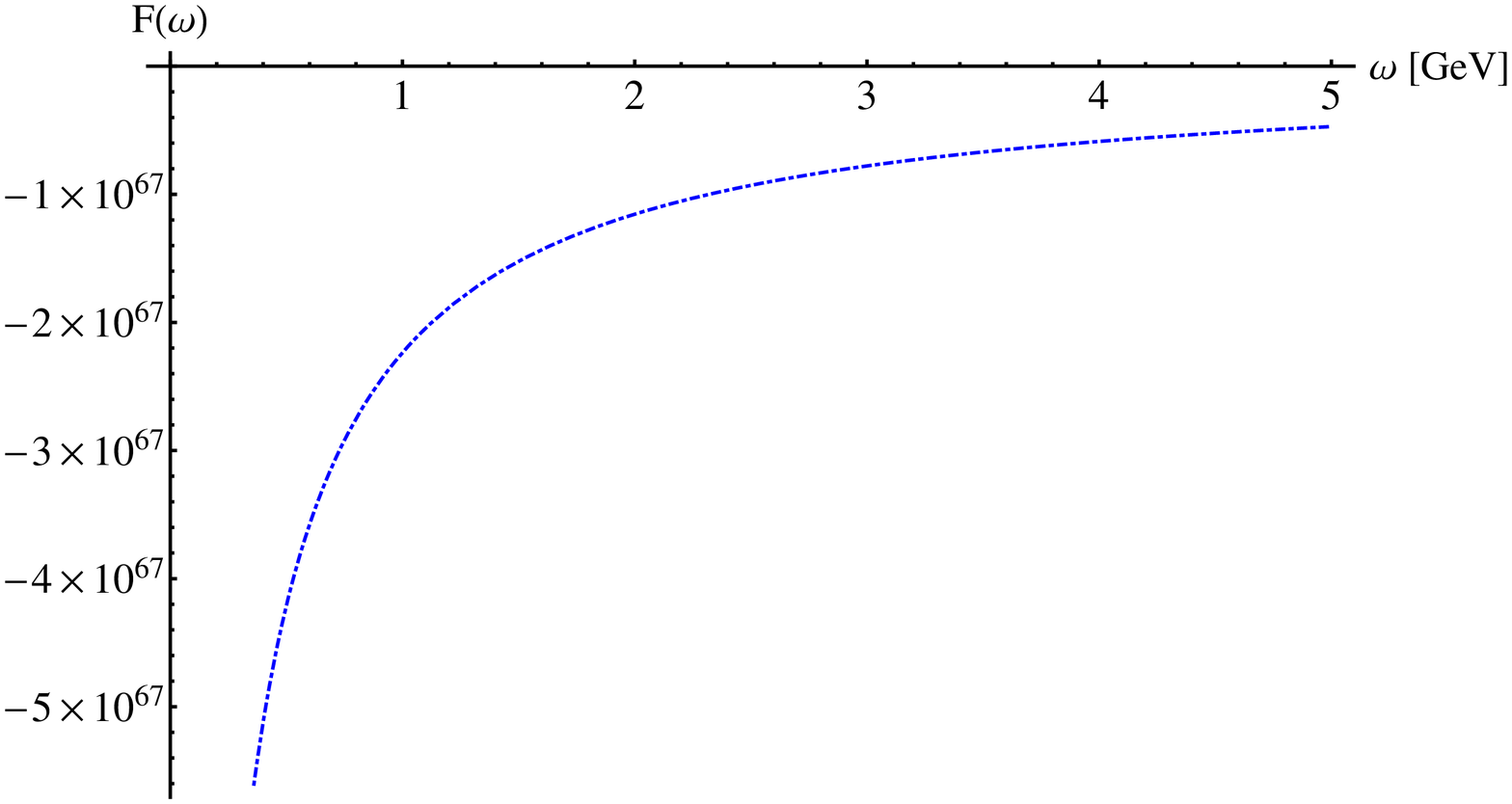}
\includegraphics[width=5.5cm]{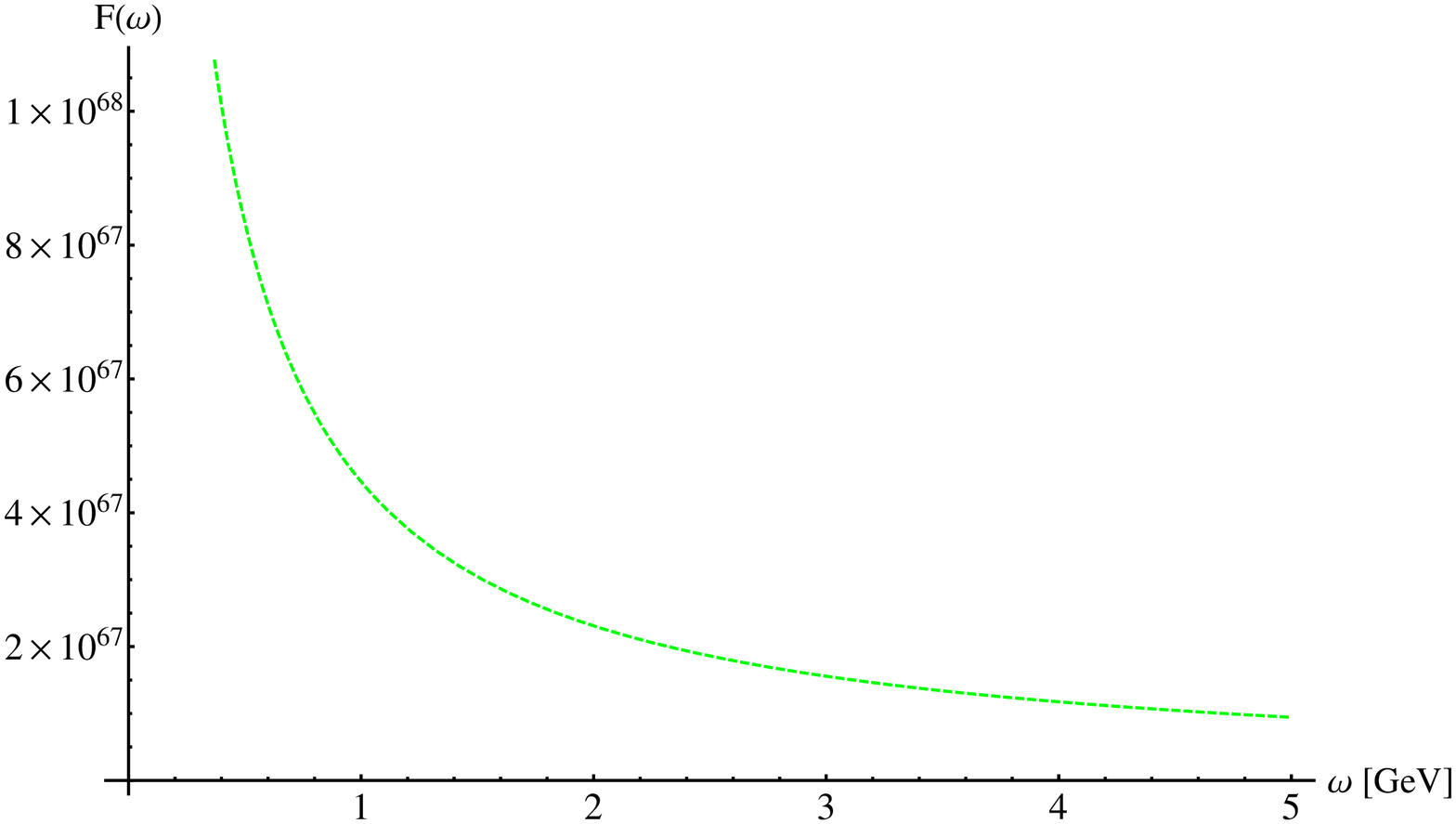}
\caption{In  natural units, the free energy of the quantum states in the inflation era is given in dependence on the energy of the scalar field $\omega$. As in Figs. \ref{fig:Nn1} and \ref{fig:Ss1}, the left and middle panels show the results at the higher and lower bounds of $\alpha$, respectively. The difference between upper- and lower-bound-results is shown in the right panel.}
\label{fig:Fe1}
\end{figure}

Left and middle panels in Fig. \ref{fig:Ss1} show the results at upper  and lower bounds of $\alpha$, respectively. Only real values are taken into consideration. The absolute values in the latter case are nearly three orders of magnitude larger than that in the earlier case.
In the earlier case $s(\omega)$ diverges at small $\omega$. It shows a kind of saturated plateau up to $\omega \sim 2$ GeV. This is almost the same as it will be shown in Fig. \ref{fig:Fe1}. One of the apparent differences is the sign. Here, $s(\omega)$ is negative. With increasing $\omega$, it decreases almost exponentially and flips its sign to negative one.
At lower bound of $\alpha$, we find that $s(\omega)$ behaves almost contrarily. We notice that $s(\omega)$ remains positive and it decreases almost exponentially. The difference between upper- and lower-bound-results is shown in the right panel. This difference is assumed to approximately give a qualitative estimation for the effects of the quantum gravity on the entropy density. It is obvious that $s(\omega)$ remains negative. Increasing $\omega$ results in decrease in the absolute values of the entropy density. It is apparent that negative entropy contradicts the laws of thermodynamics.

Also, the free energy can be deduced, directly, from  the number of quantum states, Eq. (\ref{eq:noemaga1}),
{\small
\begin{eqnarray}
F(\omega) &=& -\int  \frac{n(\omega)}{\exp(\beta \omega)-1} d \omega \nonumber \\
 &=& -\frac{1}{18 H^3 R^2 \alpha ^6 \beta } \left\{\frac{1}{\omega ^3}\left[ -12 \mathtt{atanh}(\alpha) \left(f \left(5+\alpha ^2 \left(\mu ^2-3 \omega ^2\right)\right)+6 H^2 R^2 \alpha^2 \omega ^2 \ln(R)\right)+\right.\right. \nonumber \\
&& \hspace*{7mm} \left(-9+132 \omega +2 \alpha  \left(240+36 \omega +\alpha  \left(-6 \mu ^2 \omega -3 \left(2+\mu ^2\right) \omega ^3+12 \alpha  \left(41 \omega
^3+\mu ^2 (37+41 \omega )\right) - \right.\right.\right. \nonumber \\
&& \hspace*{9mm} \left.\left.\left. 4 \alpha ^3 \left(\omega ^4 (51+32 \omega )-2 \mu ^2 \omega ^3 \left(96+\omega ^2\right)+\mu ^4 \left(-51-96 \omega
+6 \omega ^3\right)\right)\right)\right)+\right. \nonumber \\
&& \hspace*{10mm} \left.\left. 48 \alpha ^3 \left(37+34 \alpha ^2 \mu ^2\right) \omega ^2  \ln(\omega)\right)\right]+ \nonumber \\
&& \frac{1}{f^2}\left(\frac{f^3}{\omega^3} \left(47+\alpha  \left(60+2 \alpha  (11+86 \alpha ) \mu ^2+\alpha ^3 (-1+164 \alpha) \mu ^4+52 \alpha ^6 \mu ^6\right)\right) - \right. \nonumber \\
&& \hspace*{7mm} \left. \frac{1}{\omega} 6 f \alpha ^2 \left(6 H^4 R^4 \alpha ^3 \left(11+9 \alpha ^2 \mu ^2\right)+f^2 \left(11+\alpha  \left(86+\alpha  (-1+164 \alpha
) \mu ^2+78 \alpha ^4 \mu ^4\right)\right)\right)-\right. \nonumber \\
&& \hspace*{8mm} 3 \alpha ^4 \left(18 H^6 R^6 \alpha ^3+f^3 \left(-1+164 \alpha +156 \alpha ^3 \mu ^2\right)-12 f H^4 R^4 \alpha  \left(11+3 \alpha ^2 \left(-3+4
\mu ^2\right)\right)\right) \omega -  \nonumber \\
&& \hspace*{7mm} 6 H^4 R^4 \alpha ^7 \left(9 H^2 R^2-2 f \left(-16+\mu ^2\right)\right) \omega ^2+ \nonumber \\
&& \hspace*{7mm}  2 \left(26 f^3-72 f H^4 R^4-9 H^6 R^6\right) \alpha ^7 \omega ^3+ \nonumber \\
&& \hspace*{7mm}  \frac{H R \alpha ^2}{\left(f+H^2 R^2\right) \omega ^2} K \left(6 H^6 R^6 \alpha ^5 \omega ^2 (18+\omega  (9+2 \omega )) + \right. \nonumber \\
&& \left. \hspace*{7mm} f^3 \left(-9+480 \alpha +888 \alpha ^3 \left(\mu ^2+2 \omega ^2\right)+136 \alpha ^5 \left(3 \mu ^4+14 \mu ^2 \omega ^2-2
\omega ^4\right)\right)+\right. \nonumber \\
&& \hspace*{7mm} f H^4 R^4 \alpha ^3 \left(24 (33-4 \omega ) \omega +\alpha ^2 \left(\mu ^2 \left(27+702 \omega -270 \omega ^2-8 \omega ^4\right)+\omega ^2 (162+\omega
 (405+98 \omega ))\right)\right)+ \nonumber \\
&& \hspace*{7mm} \left. f^2 H^2 R^2 \left(-9+480 \alpha +24 \alpha ^3 \left(37 \mu ^2+\omega  (33+70 \omega )\right) + \right.\right. \nonumber \\
&& \hspace*{7mm} \left.\left. \alpha ^5 \left(3 \omega ^2 (18+(117-62 \omega
) \omega )+8 \mu ^4 \left(51+\omega ^2\right)+\mu ^2 \left(27+2 \omega  \left(351+644 \omega -4 \omega ^3\right)\right)\right)\right)\right)+ \nonumber \\
&& \hspace*{7mm} 12 H^4 R^4 \alpha ^5 \left(-3 H^2 R^2 \alpha ^2+8 f \left(1+4 \alpha ^2 \mu ^2\right)\right)\ln(\omega)+ \nonumber \\
&& \hspace*{7mm} \frac{3}{\mu } i \sqrt{f} H R \alpha ^2 \left[f \left(f+H^2 R^2\right) (-3+160 \alpha )+3 \alpha ^3 \left(296 f^2+15 H^4 R^4 \alpha ^2+f H^2
R^2 \left(88+9 \alpha ^2\right)\right) \mu ^2 + \right. \nonumber \\
&& \hspace*{12mm} \left. 10 f \left(68 f+5 H^2 R^2\right) \alpha ^5 \mu ^4\right]  \ln\left[\left(2 i \sqrt{f} \mu -2 K\right)/ \right. \nonumber \\
&& \hspace*{7mm} \left. \left(f H R \alpha ^2 \left(f \left(f+H^2
R^2\right) (-3+160 \alpha )+3 \alpha ^3 \left(296 f^2+15 H^4 R^4 \alpha ^2+f H^2 R^2 \left(88+9 \alpha ^2\right)\right) \mu ^2+\right. \right. \right.\nonumber \\
&&  \hspace*{12mm} \left.  \left. \left.10 f \left(68 f+5
H^2 R^2\right) \alpha ^5 \mu ^4\right) \omega )]- \right.\right. \nonumber \\
&& \hspace*{7mm}  \left.\left.\frac{9 f H^3 R^3 \alpha ^5 }{\sqrt{f+H^2 R^2}}\left[ 88 \left(f+H^2 R^2\right)+3 \left(39 f+28 H^2 R^2\right) \alpha
^2 \mu ^2\right]  \ln \left[2 \left(f \omega +H^2 R^2 \omega +\sqrt{f+H^2 R^2} K \right)\right]\right)\right\}. \nonumber
\end{eqnarray}
}
Again, we notice that the expression contains complex terms. The results at the upper and lower bounds of $\alpha$ are given in Fig. \ref{fig:Fe1}: left and middle panel, respectively. The difference between $s(\omega)$ at upper and lower bound of $\alpha$ is given in the right panel.

The dependence of free energy on $\omega$ is illustrated in Fig. \ref{fig:Fe1}. In the left panel, we show the results at the upper bound of $\alpha$. In doing this, we take into consideration the real values, only. We notice that the free energy diverges to negative values at very small values of $\omega$. Then, $F(\omega)$ makes a plateau up to $\omega\sim 2$ GeV. With increasing $\omega$, the free energy arising from the quantum states switches to positive values. Afterwards, it increase, nearly exponentially. The middle panel shows the result at the lower bound of $\alpha$. We notice that the absolute value of  $\alpha$ is about three orders of magnitude larger than in the case of upper bound (left panel). Also, we notice that lower-bound-values remain negative although they exponentially decay with increasing $\omega$. The difference between upper- and lower-bound-results is shown in the right panel. It give a qualitative estimation for the effects of the quantum gravity i.e., GUP on the free energy when taking into consideration the quantum gravity i.e., applying the GUP approach.

\end{document}